\documentclass[review,12pt]{elsarticle}
\usepackage{lineno,hyperref}
\modulolinenumbers[5]
\usepackage{amsmath}
\usepackage{amsfonts}
\usepackage{amssymb}
\usepackage{cleveref}
\usepackage{lmodern}
\usepackage{xfrac}
\usepackage{graphicx}
\usepackage{subfig}
\graphicspath{ {./figures/} }
\usepackage{fullpage}
\usepackage{float}
\usepackage{caption}
\usepackage{siunitx}
\usepackage{bm}

\usepackage{xcolor}
\DeclareMathOperator*{\argmin}{arg\,min}

\biboptions{sort&compress}

\usepackage{lipsum}
\makeatletter
\def\ps@pprintTitle{%
 \let\@oddhead\@empty
 \let\@evenhead\@empty
 \def\@oddfoot{}%
 \let\@evenfoot\@oddfoot}
\makeatother

\begin{document}

\begin{frontmatter}

\title{Uncertainty Analysis of Microsegregation during Laser~Powder~Bed~Fusion}
\author[1]{Supriyo Ghosh\corref{mycorrespondingauthor}}
\cortext[mycorrespondingauthor]{Corresponding author.}
\ead{gsupriyo2004@gmail.com}
\author[2]{Mohamad~Mahmoudi}
\author[1]{Luke~Johnson}
\author[1,2]{Alaa~Elwany}
\author[1,2,3]{Raymundo~Arroyave}
\author[3]{Douglas~Allaire}
\address[1]{Department of Materials Science \& Engineering, Texas A\&M University, College Station, TX 77843, USA}
\address[2]{Department of Industrial \& Systems Engineering, Texas A\&M University, College Station, TX 77843, USA}
\address[3]{Department of Mechanical Engineering, Texas A\&M University, College Station, TX 77843, USA}
\begin{abstract}
Quality control in additive manufacturing can be achieved through variation control of the quantity of interest (QoI). We choose in this work the microstructural microsegregation to be our QoI. Microsegregation results from the spatial redistribution of a solute element across the solid-liquid interface that forms during solidification of an alloy melt pool during the laser powder bed fusion process. Since the process as well as the alloy parameters contribute to the statistical variation in microstructural features, uncertainty analysis of the QoI is essential. High-throughput phase-field simulations estimate the solid-liquid interfaces that grow for the melt pool solidification conditions that were estimated from finite element simulations. Microsegregation was determined from the simulated interfaces for different process and alloy parameters. Correlation, regression, and surrogate model analyses were used to quantify the contribution of different sources of uncertainty to the QoI variability. We found negligible contributions of thermal gradient and Gibbs-Thomson coefficient and considerable contributions of solidification velocity, liquid diffusivity, and segregation coefficient on the QoI. Cumulative distribution functions and probability density functions were used to analyze the distribution of the QoI during solidification. Our approach, for the first time, identifies the uncertainty sources and frequency densities of the QoI in the solidification regime relevant to additive manufacturing.
\end{abstract}

\begin{keyword}
Additive manufacturing \sep phase-field modeling \sep finite element analysis  \sep surrogate modeling \sep dendrite \sep microsegregation
\end{keyword}

\end{frontmatter}
\section{Introduction}
Additive manufacturing (AM) processes are increasingly becoming pervasive due to their ability to produce intricate parts with improved properties compared to traditional manufacturing processes and are therefore widely being used in aerospace and automotive industries~\cite{frazier2014,Herzog2016}. The AM process studied in the present work is laser powder bed fusion (LPBF). The material chosen for research is a Ni-Nb alloy, which is a binary analog of Inconel 718, that finds application in jet-engine and gas-turbine components~\cite{frazier2014,Herzog2016}. During the LPBF process, the laser beam moves across the alloy powder with a fixed velocity, resulting in a liquid melt pool that solidifies into a complex solid-liquid interface. The morphology of the solid-liquid interface is determined by the solidification conditions in the melt pool, namely, temperature gradient $G$ and solidification velocity $V$ \cite{supriyo2017}. The typical solidification morphologies are planar, cellular, and dendritic in which the key microstructural features such as spacing, segregation, and orientation determine the properties of an AM component.

\begin{figure}[h]
\begin{center}
\includegraphics[scale=0.5]{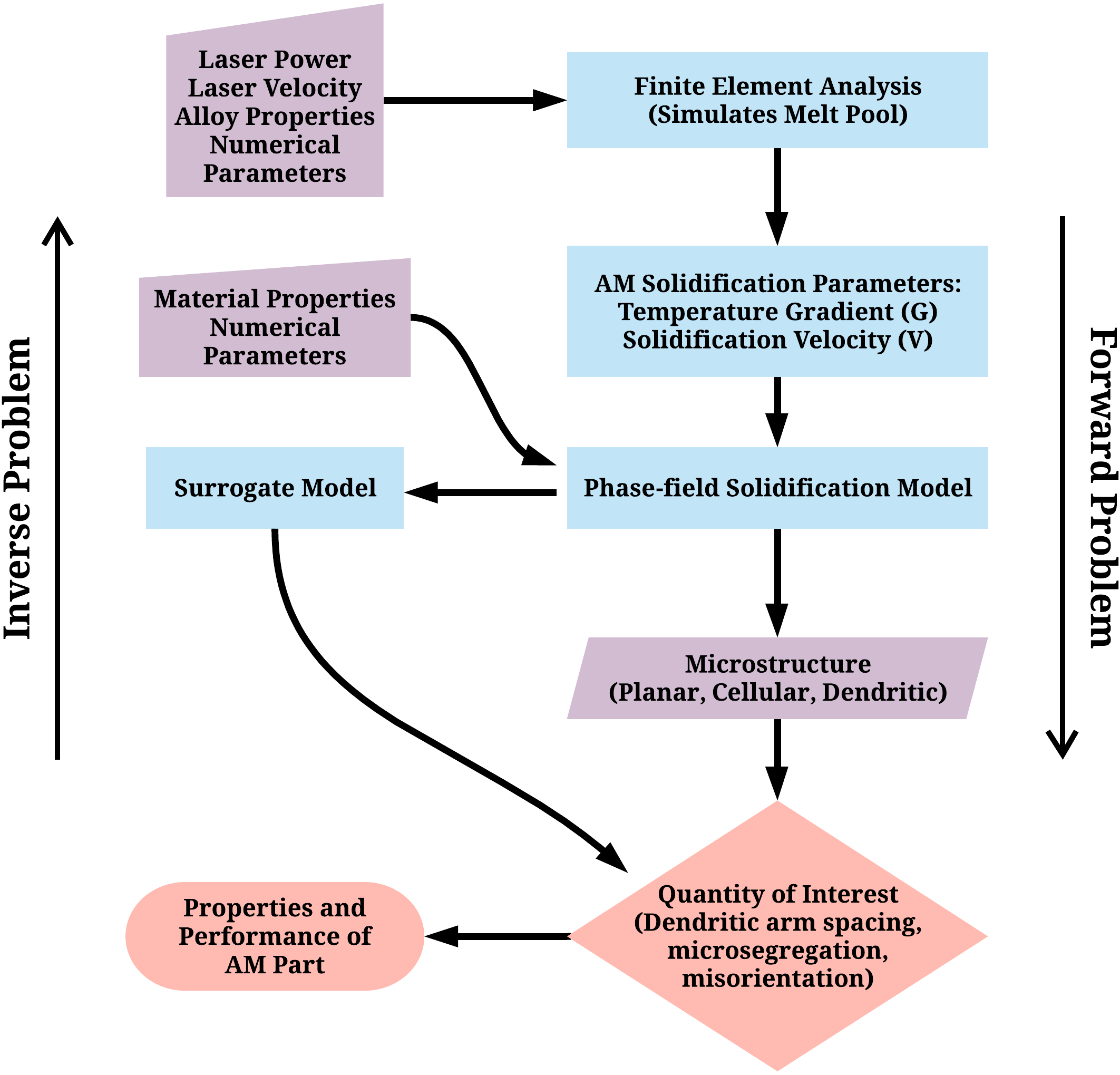}
\caption{Summary flowchart of uncertainty propagation in a multi-level simulation framework.}\label{fig_schematic}
\end{center}
\end{figure}

The primary barrier to the widespread adoption of the new classes of AM materials is the lack of confidence in the product quality~\cite{Khairallah2016,King2014,King2015}, which is due to the variability present at various stages of the manufacturing process. Development of a quality control approach based on uncertainty quantification (UQ) of the process through multiscale simulations is a key to resolve this issue. The flowchart in Fig.~\ref{fig_schematic} summarizes our multiscale uncertainty analysis approach. A macro-scale finite element analysis (FEA) model simulates the LPBF process where the potential sources of uncertainty are the laser processing parameters, material properties, and boundary conditions. The melt pool profile obtained from FEA simulations is used to determine the solidification parameters as they are inputs to the microstructure model. A phase-field model~\cite{boettinger2002,chen2002,steinbach2009} is used for the simulation of microstructure, from which the quantity of interests (QoIs) are determined. Phase-field models~\cite{attari2018,ghosh2017_spinodal,tamoghna2018} can be computationally intensive as the simulations often run from hours to days to produce informative microstructures. Surrogate models are inexpensive approximations to the original computer model (Ref.~\cite{allaire2010} and the references within). Therefore, a suitable surrogate model could potentially substitute for the expensive phase-field simulations.

The microstructural features that develop during LPBF solidification process are statistically variable in several aspects~\cite{Franco2017_sr,ghosh_review}. The solidification morphology and the distribution of the grain size can vary from region to region within the microstructure. The spatial distribution of the solute composition in the as-built alloy samples can be severely inhomogeneous due to rapid cooling rates ($\dot{T} = G \times V$). The misorientation of the solidifying grains with respect to the build direction is frequently observed in multi-layer multi-pass solidification microstructures~\cite{debroy2015}. The above microstructural features further vary as a function of the alloy properties, such as the segregation coefficient, the diffusivity of the liquid, and the Gibbs-Thomson coefficient. All the above issues lead to inhomogeneous mechanical properties of the solid material. From a modeling perspective, numerical parameters such as interface width and mesh size contribute to uncertainties. It is entirely unavoidable to eliminate the microstructural variabilities due to the non-equilibrium nature of the solidification during LPBF~\cite{King2015,Khairallah2016}. The incorporation of non-equilibrium physics adds further computational cost to the original computer simulation models, thus requiring efficient surrogate models to approximate those.

We chose to study the microsegregation (i.e., the spatial distribution of the solute across the interface) in LPBF solidification microstructures as our QoI. Microsegregation severely affects the yield, tensile, and fatigue strengths of the material and can be estimated from a microstructure in a relatively small domain. Our strategy in this work is to use a microstructure model to determine the QoI statistics in LPBF microstructures and to apply the knowledge of uncertainty quantification to understand the contribution of various sources to the QoI variability. Microstructure evolution during a low-velocity solidification process possesses much less variability of the microstructural features compared to that of the same during the LPBF process and hence is often ignored in the casting solidification literature. Conversely, uncertainty analysis is essential for the quantification of LPBF microstructures due to significant variability in its key features. It is, however, rarely addressed in the literature (for recent overviews, refer to~\cite{mahadevan2018,mahadevan2017,brandon2016}). 

Uncertainty quantification frameworks of a FEA model for melt pool simulation and a cellular automata model for microstructure simulation were coupled very recently~\cite{mahadevan2018,claire2018}. The phase-field method is more efficient compared to the cellular automata method in that the phase-field method does not require explicit tracking of the solid-liquid interface and the difficulties associated with the estimation of interfacial curvature are handled efficiently~\cite{zaeem2013,rappaz2016}. There is no study, as far as we are aware, that addresses uncertainty during the high-fidelity microstructure evolution using the phase-field method. Further, one must be cautious in using statistics derived from limited input and output data during phase-field simulation of the LPBF process, as small datasets may not necessarily represent the overall processing as well as the solidification map. Therefore, the distribution derived from a large dataset could potentially be used as a metric to gain a deeper understanding of the physics and trends involved in the solidification process. Uncertainty analysis for LPBF is a relatively new research field with tremendous growth opportunities in the future. In this first approach, key questions we address in the microsegregation modeling in the solidification regime relevant to LPBF are: Which model parameter uncertainties affect microsegregation? How can statistical distributions be used to represent microsegregation? How do sample size and distribution affect microsegregation? Are surrogate models good enough to approximate phase-field models?

The remainder of the article is as follows. Simulation method, parameters, and microstructure analysis techniques are outlined in Sec.~\ref{sec_details}. The sensitivity, surrogate, and frequency analyses of our phase-field simulation results are illustrated in Sec.~\ref{sec_results}. A detailed discussion of our simulation results is given in Sec.~\ref{sec_discussion}. Finally, a summary and outlook of the current work are given in Sec.~\ref{sec_summary}.

\section{Simulation Details}\label{sec_details}
\subsection{Method}\label{sec_method}
Inconel 718 (IN718) is a multicomponent alloy with a more than ten alloying elements making the microsegregation in the solidifying melt pool complex. Our strategy here is to conduct an uncertainty analysis of a binary approximation of the IN718 alloy, the understanding of which can be effectively extended for the modeling of a multicomponent analog using a multi-phase-field model that handles multiple components and phases~\cite{supriyo20173d,kundin2015,steinbach2012}. Ni-Nb is the most important binary analog of IN718, since Nb segregates most severely from the solid to the liquid during directional solidification of the IN718 melt pool due to the smallest partition coefficient of Nb among all elements in IN718 and thus has a major role in determining the key solidification microstructural features as well as the subsequent solid-solid precipitation reactions.

We have used a phase-field solidification model detailed in Refs.~\cite{Echebarria2004,Karma2001} to simulate the solid-liquid interfaces that evolve during directional solidification of a Ni-5~wt\% Nb alloy. The model was developed to obtain a quantitative assessment of the microstructure information that results during directional growth of a binary alloy. The model is employed and validated extensively in the literature~\cite{Echebarria2004,steinbach2009,provatas2011} and the simulation results agree excellently with the experimental measurements on metallic alloys in the low-velocity limit. Therefore, as a reference, such an established model is programmed to simulate the evolution of the non-conserved phase-field $\phi$ and the conserved composition field $c$ during the time($t$)-dependent solidification process in the melt pool. The phase-field $\phi$ is a scalar-valued order parameter field which distinguishes the microstructure phases; $\phi$ = 1 in the solid, $\phi$ = -1 in the liquid and the solid-liquid interface is described by $-1<\phi<1$. This approach avoids explicit tracking of the interface and thus the complex solid-liquid interfaces are extracted in an efficient way. The effects of melt convection are not included, the diffusion of solute in the solid is neglected, the diffusion of heat is ignored, and local equilibrium at the solid-liquid interface is imposed in this model. The evolution equation for $\phi$ can be written as:

\begin{eqnarray}\label{eq_phi}
\tau_0 a(\hat{q})^2\frac{\partial \phi}{\partial t} = W_{0}^{2} \nabla \cdot \left[{a(\hat{q})}^2 \nabla\phi\right] + \sum_{i=1}^{d} \partial_i \left[a(\hat{q}) \frac{\partial a(\hat{q})}{\partial(\partial_i \phi)} |\nabla\phi |^2 \right] \nonumber \\
+ \phi -\phi^3 -  \frac{\lambda}{1-k} (1-\phi^2)^2 \left[\exp(u) -1 + \frac{G(z-Vt)}{m_l c_0/k}\right].
\end{eqnarray}
The dimensionless surface energy function $a(\hat{q}) = 1 - \epsilon \left[3- 4 \sum_{i=1}^{d} q_{i}^{4}\right]$ represents the $d$-dimensional fourfold anisotropy at the solid-liquid interface with a strength $\epsilon$ and $q_i$ is the interface normal vector pointing into liquid along the Cartesian direction $i$ in the lab frame of reference. Alloy composition $c_0$, liquidus slope $m_l$, and equilibrium partition coefficient (or, segregation coefficient) $k$ can be approximated from a Ni-Nb phase diagram~\cite{baker1992,knorovsky1989}. The dimensionless chemical potential $u$ is given by $\ln \left(\frac{2ck/c_0}{1+k-(1-k)\phi}\right)$. A frozen-temperature approximation~\cite{Echebarria2004} is applied in which the temperature gradient $G$ is translated along the $z$ (growth) axis with a velocity $V$.

The evolution equation for $c$ can be written as:
\begin{equation}\label{eq_c}
\frac{\partial c}{\partial t} = -\nabla \cdot \left[ - \frac{1}{2}(1+\phi)\, D \, c \, \exp(u)^{-1}  \, \nabla\exp(u) + a_t W_0 (1-k) \exp(u) \frac{\partial \phi}{\partial t} \frac{\nabla\phi}{|\nabla \phi|}\right],
\end{equation}
where the first term inside the square bracket represents a standard Fickian diffusion flux, and the second term is the anti-trapping solute flux term that prevents any artificial solute trapping to occur at the simulated solid-liquid interface and thus the solute redistribution across the interface becomes efficient. The value of $a_t = 1/(2\sqrt{2})$ is based on the thin interface model by Karma~\cite{Karma2001}. $D$ is the diffusivity of solute in the liquid.

The numerical parameters in the model, the interface thickness $W_0$, the phase-field relaxation time $\tau_0$, and the dimensionless coupling constant $\lambda$, are linked to material parameters via the chemical capillary length $d_0 = 0.8839 W_{0}/\lambda$ and the timescale for diffusion $\tau_0 = 0.6267\lambda W_{0}^{2}/D$ that use a thin-interface analysis in order to make the interface kinetics vanish. Both $W_0$ and $\tau_0$ values are used to render all the simulation parameters dimensionless.
\subsection{Parameters and procedures}\label{sec_parameters}
\subsection{Finite element analysis}
\begin{figure}
\centering
\includegraphics[scale=0.8]{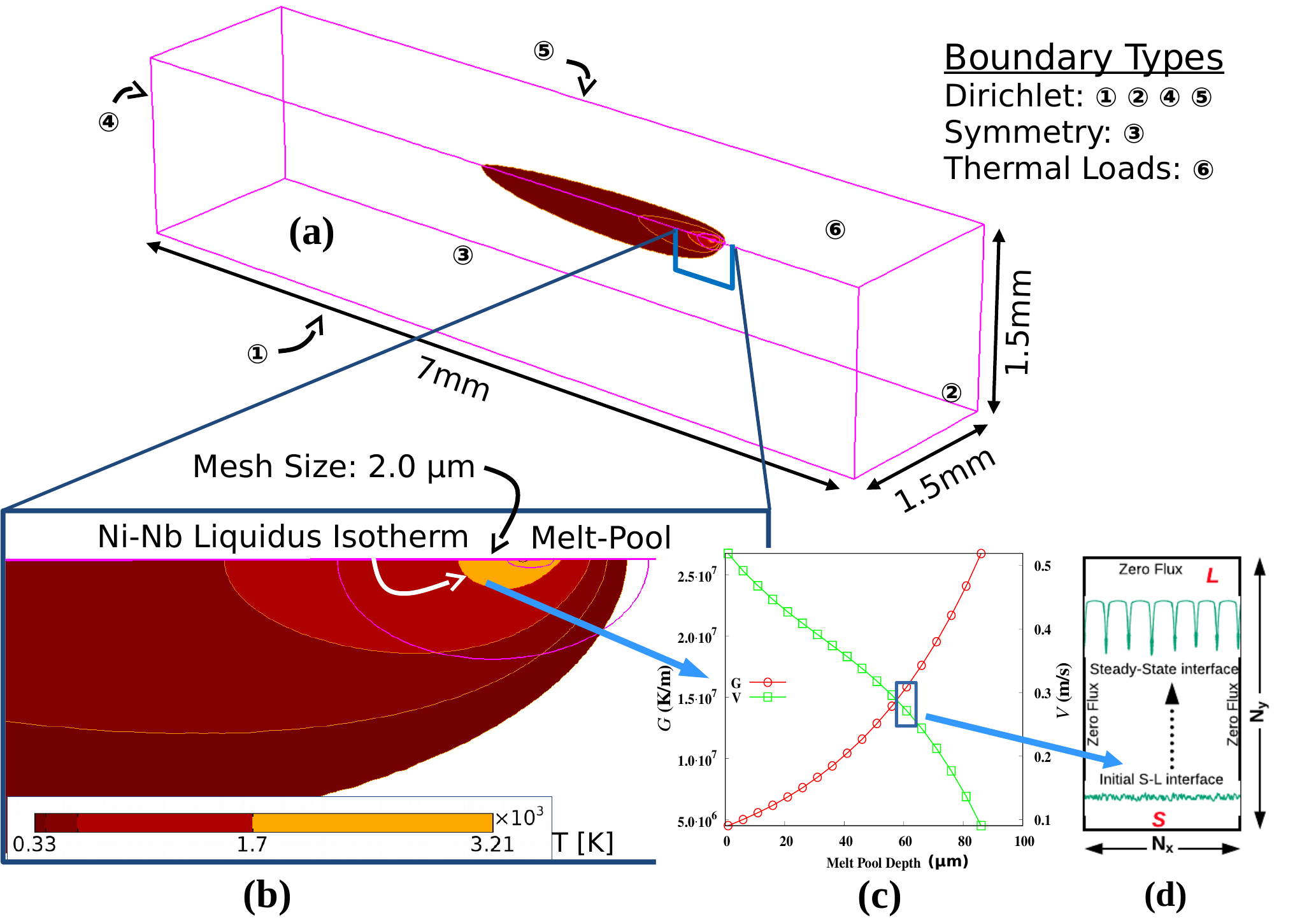}
\caption{(a) Three-dimensional FEA simulations determine the temperature distribution in the resulting melt pool during LPBF (laser power of 200 W and beam speed of 500 mm s$^{-1}$) of a Ni-5\% Nb alloy. A typical melt pool profile is shown within the geometry of the simulation domain. (b) A 2D section of the 3D melt pool cut along its center in the length direction is presented to illustrate the temperature distribution. (c) $G$ and $V$ vary as a function of the melt pool depth. (d) These $G$ and $V$ values are the inputs to the phase-field simulation. Note that this schematic is one instance of multiple such parallel FEA and phase-field calculations that are being used for the uncertainty analysis of the LPBF microstructures.}\label{fig_process}
\end{figure}

The solidification parameters in Eq.~(\ref{eq_phi}) were estimated using 3D finite element analysis simulations carried out within the COMSOL Multiphysics~\cite{Comsol} heat transfer module. A single-track laser scan in the length direction of a rectangular parallelepiped specimen of Ni-Nb substrate at an initial uniform temperature of 298 K was modeled (Fig.~\ref{fig_process}). The laser beam power distribution was assumed to be Gaussian with a beam diameter of \SI{80}{\micro \meter}. The power absorption coefficient was considered to be phase-dependent following Ref.~\cite{trapp2017}. The absorptivity values of solid, liquid, and vapor phases was 0.3, 0.3, and 0.6, respectively. The governing equations for the conservation of energy during the laser heat distribution within the material can be found in Ref.~\cite{karayagiz2018}. The boundary conditions listed as thermal loads (Fig.~\ref{fig_process}a) included heat transfer considerations of the deposited laser beam, natural convection, radiation, and vaporization. Natural convection and radiation contributions were included through the standard implementations of each phenomenon typical of heat transfer analysis~\cite{karayagiz2018}. Vaporization was included through a loss of energy via the mass flux of vapor leaving the system which was calculated using the weld pool evaporative flux model developed in Ref.~\cite{bolten1984}.

Heat transfer within the simulation domain was governed by conduction only, but included phase-transformations effects in the form of latent heat contributions to the specific heat during melting and vaporization, as well as phase-dependent material properties~\cite{karayagiz2018}. Thermophysical properties such as the bulk material density, latent heat, and specific heat of the solid, liquid, and vapor phases of the Ni-Nb alloy were calculated using an atomic composition based rule of mixtures of pure Ni and Nb properties. For a small alloying addition of Nb (5 wt\%), the properties are essentially that of pure Ni which can be found in the material property handbook~\cite{asmnickel,agarwal2004}. Liquid conductivity was calculated as twice that of the solid to approximate the effect of convection within the melt pool, as is common practice in heat transfer analysis~\cite{ladani2017}. Vapor phase conductivity was artificially enhanced to simulate the transmission of laser deposited energy through the phase into the substrate during vaporization. This enhanced vapor conductivity is unique to this model, but has similarities to the element birth and death method~\cite{roberts2009}. However, our approach is more physical in that it retains the vapor that forms in the region between the laser beam and the substrate material.

The LPBF process was simulated for a wide range of laser parameters that were based on our knowledge and machine specifications; laser power typically varied between 30 W and 300 W and beam speed typically varied between 0.1 m s$^{-1}$ and 2.5 m s$^{-1}$. A typical resultant thermal profile after the FEA simulation is shown in Fig.~\ref{fig_process}b. The melt pool profile obtained from this simulation was used to extract the solidification parameters $G$ and $V$ (following their mathematical expressions given in Refs.~\cite{supriyo2017,raghavan2016}) at the liquidus temperature isotherm where the solid-liquid phase transformation begins. The ranges of $G$ and $V$ (Fig.~\ref{fig_process}c) estimated for various values of laser parameters approximate the entire LPBF solidification map (Table~\ref{table_param}). These $G$ and $V$ values are inputs to the phase-field model (Fig.~\ref{fig_process}d).

It should be noted that we ignore the effects of uncertainty regarding the material properties (e.g., laser absorptivity, thermal conductivity of the liquid), melt pool physics (e.g., computational fluid dynamics and Marangoni convection), and numerical parameters (e.g., mesh size) that affect the temperature distribution as well as the melt pool size and geometry during our FEA simulations; a detailed uncertainty analysis of which can be found in Ref.~\cite{mahmoudi2018}. An arbitrary single set of laser parameter may not necessarily leads to ranges of G and V that approximate the overall LPBF process and solidification maps, on average. This is why we choose to use wide ranges of laser parameters in our FEA simulations to use the resultant ranges of G and V values for the subsequent phase-field simulations. A large number of input samples designed from the resultant ranges of $G$ and $V$ is used to perform uncertainty analysis efficiently in the present work.

\subsection{Phase-field}
Equations~(\ref{eq_phi}) and~(\ref{eq_c}) are solved on a two-dimensional uniform mesh using a finite volume method and no-flux boundary conditions in all directions. The numerical values of the parameters used in the simulations are: interface width of 0.5 nm; grid spacing of 0.3 nm; and two different mesh sizes (in grid units) of 600($N_x$)$\times$3000($N_y$) and 10($N_x$)$\times$3000($N_y$) are used to simulate cellular and/or planar interfaces. The interface thickness and grid spacing values used in our simulations are small enough so that the results become independent of their values. Each simulation begins with a thin layer of solid at the bottom of the simulation box with an initial Nb composition of $kc_0$ in the solid and $c_0$ in the liquid. Small, random amplitude perturbations are applied at the initial solid-liquid interface, from which perturbations either grow with time and break into a cellular interface or decay with time resulting into a planar interface. The spatial redistribution of the resulting composition field $c$ across the interface is referred to as microsegregation. 

Three input material parameters, namely, the diffusivity of the liquid $D$, the Gibbs-Thomson coefficient $\Gamma$, and the segregation coefficient $k$ are found to affect significantly on the variability of the phase-field model output. Gibbs-Thomson coefficient $\Gamma$ is given by $d_0 \, \Delta T_0$, where $d_0$ is the capillary length and $\Delta T_0$ is the equilibrium freezing range of the alloy. Although not shown here, we found that the variabilities of $c_0$, $m_l$, and $\epsilon$ (refer to Eq.~(\ref{eq_phi})) are not significant and hence are ignored following preliminary phase-field experiments. The specific ranges of the uncertain material properties are decided based on the literature review~\cite{baker1992,lee2010,knorovsky1989,asmnickel,agarwal2004}. 

We use a Latin Hypercube Sampling (LHS) method~\cite{olsson2003latin} to sample $N = 100$ process and alloy parameters evenly across all possible values, the statistical measures of which are listed in Table~\ref{table_param}. LHS partitions each input distribution into $N$ intervals of equal probability and select one sample randomly from each interval to construct the input design space. This is more efficient compared to the Monte Carlo Sampling (MCS) technique that randomly selects a sample from an input distribution, leading to some intervals in the sample space with a clustered data and some intervals with no samples. 

\begin{table}[h]
\begin{center}
\begin{tabular}{l  c c c  c}
\hline \hline
Input ($\bm{X}$) & Minimum & Maximum \\ \hline
$G$ (K m$^{-1}$)& $10^6$ & $4\times 10^7$  \\ 
$V$ (m s$^{-1}$) & 0.01  & 2.5   \\ 
$D$ (m$^2$ s$^{-1}$) & $10^{-9}$ & $9 \times 10^{-9}$  \\ 
$\Gamma$ (K m) & $10^{-7}$ & $4 \times 10^{-7}$  \\ 
$k$ (dimensionless) & 0.48 & 0.74  \\ 
\hline \hline
\end{tabular}
\caption{Assumed ranges for the phase-field model input parameters.}\label{table_param}
\end{center}
\end{table}

\subsection{Analysis of uncertainty}\label{sec_analysis}
We use a scatterplot to visualize the variability of $n$ input process and material parameters $\bm{X} = \lbrace\bm{x}_{i1},\ldots,\bm{x}_{in}\rbrace_{i=1}^{N}$ on the QoI $\bm{Y}$. We perform sensitivity analysis by determining the Pearson product-moment correlation coefficient $\rho_{\bm{X}\bm{Y}}$ for each $\bm{X}$ and $\bm{Y}$ combinations~\cite{montgomery2014,devore1987}:
\begin{equation}
\text{Corr}\,(\bm{X}, \bm{Y}) = \rho_{\bm{X}\bm{Y}} = \frac{\text{Covariance}\,(\bm{X},\bm{Y})}{\sigma_{\bm{X}} \, \sigma_{\bm{Y}}},
\end{equation}
where $\sigma_{\bm{X}}$ and $\sigma_{\bm{Y}}$ are the standard deviations of $\bm{X}$ and $\bm{Y}$, respectively. The value of $\rho_{\bm{X}\bm{Y}}$ ranges between -1 and 1; $\rho_{\bm{X}\bm{Y}} = 0$ signifies no correlation between $\bm{X}$ and $\bm{Y}$, and the absolute value of $\rho_{\bm{X}\bm{Y}}$ defines the relationship strength between the two. A value of $\rho_{\bm{X}\bm{Y}}$ close to 0 signifies that the variability of the input $\bm{X}$ on $\bm{Y}$ can be discarded. From such an analysis, the number of uncertain input parameters can be reduced and the parameters can be ranked according to their correlation strengths with $\bm{Y}$. 

A regression analysis~\cite{montgomery2014,devore1987} was used to obtain the line of best fit from a typical scatterplot representing $\bm{X}$ \textit{vs.} $\bm{Y}$. Regression line equations, as a first approximation, would help to find a rough estimate of the QoI $\bm{Y}$ for arbitrary values of the input choices $\bm{X}$. The calculations of mean ($\mu$) and standard deviation ($\sigma$) of $\bm{X}$ and $\bm{Y}$ are necessary to perform regression analysis, which can be expressed as: $\bm{Y} = a\,\bm{X} + b$, where $a = \rho_{\bm{X}\bm{Y}}\, \sigma_{\bm{Y}}/\sigma_{\bm{X}}$ and $b = \mu_{\bm{Y}} - a\,\mu_{\bm{X}}$. $\sigma_{\bm{X}}$ = $\sqrt{\sum_i^N(x_i-\mu_{\bm{X}})^2/N}$ and $\sigma_{\bm{Y}}$ = $\sqrt{\sum_i^N(y_i-\mu_{\bm{Y}})^2/N}$, where $x_i$ and $y_i$ are the elements in $\bm{X}$ and $\bm{Y}$, respectively.

A Gaussian process (GP) surrogate model~\cite{kennedy2001} was used to approximate $\bm{Y}$. The surrogate model was built on the training dataset ($\bm{X}$, $\bm{Y}$) provided by the phase-field simulations. The performance of the surrogate model was estimated by measuring the difference between phase-field model ($y_i$) and surrogate model (${\hat{y}}_i$) predictions, referred to as the mean absolute predictive error (MAPE):
\begin{equation}\label{eq_mape}
{\rm MAPE} = \frac{1}{N}\sum_{i=1}^{N}\left|y_{i}-\hat{y}_{i}\right|\quad\forall i\in\left\{ 1,\ldots,N\right\}.
\end{equation}
Here $y_{i}$ is $i$-th element of the phase-field model output at an input $\bm{x}_{i}$, and $\hat{y}_{i}$ is the $i$-th element of the surrogate model prediction evaluated at the same input $\bm{x}_{i}$. Similarly, mean absolute percent error is given by $\frac{1}{N}\sum_{i=1}^{N}\frac{\left|y_{i}-\hat{y}_{i}\right|}{y_i} \times 100$. 

Frequency analysis of $\bm{Y}$ is performed using cumulative distribution (CDF) and probability density functions (PDF)~\cite{montgomery2014,devore1987}. Phase-field simulated $\bm{Y}$ values are ranked $M=1$ through $M=N=100$ in the ascending order and then, for each $\bm{Y}$ entry, the empirical CDF is calculated using the formula: $M/N$ and plotted against $\bm{Y}$. The derivative at any point in this CDF plot is regarded as the probability of the associated PDF. The empirical CDF is fitted against different target CDFs such as normal, lognormal, gamma, beta, and Weibull in a MATLAB distribution fitting tool `dfittool'~\cite{Matlab} that uses maximum likelihood estimator (MLE) to approximate $\bm{Y}$. This analysis helps to gain insights regarding the type of frequency distribution the simulated $\bm{Y}$ closely follows. Such an analysis was performed for various random samples to test whether there is any effect of sampling on the $\bm{Y}$. 

\section{Results}\label{sec_results}
\subsection{General features}\label{sec_general}
The solid-liquid interfaces in a solidifying melt pool grow in the liquid in the direction of the temperature gradient $G$ at a rate $V$. The morphology of the solidifying interface can be estimated by two critical velocities~\cite{rappazbook}, the constitutional supercooling velocity $V_{cs} = G D / \Delta T_0$ and the absolute velocity $V_{ab} = \Delta T_0 D / (k \Gamma)$. When $V$ is below $V_{cs}$, the interface morphology would be planar, for $V_{cs}<V<V_{ab}$, the interface would be cellular/dendritic, and, for $V>V_{ab}$, the interface becomes planar again. 

\begin{figure}[h]
\centering
\subfloat[]{\label{fig_planar}\includegraphics[scale=0.08]{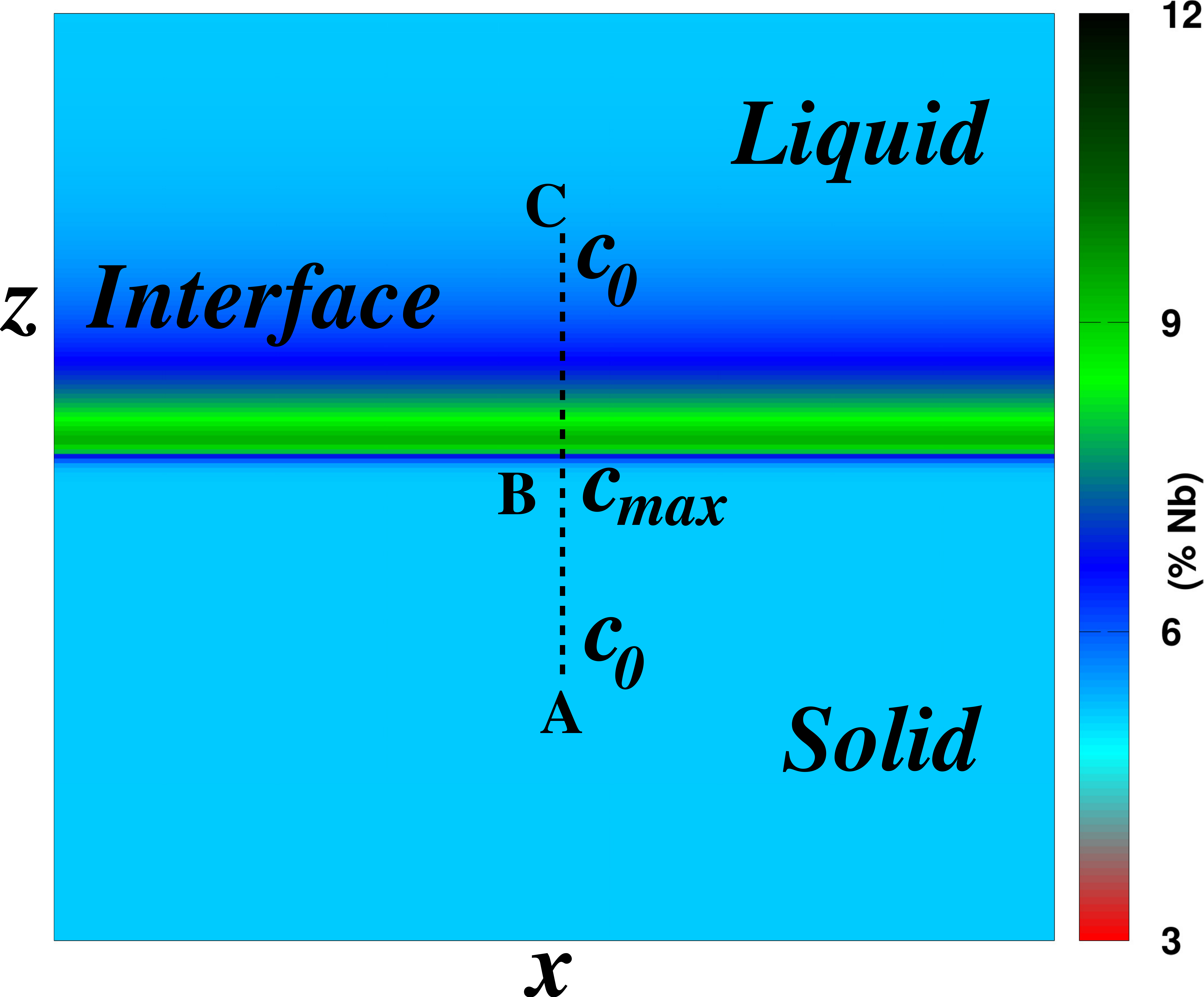}}\hspace{5mm}
\subfloat[]{\label{fig_cell}\includegraphics[scale=0.08]{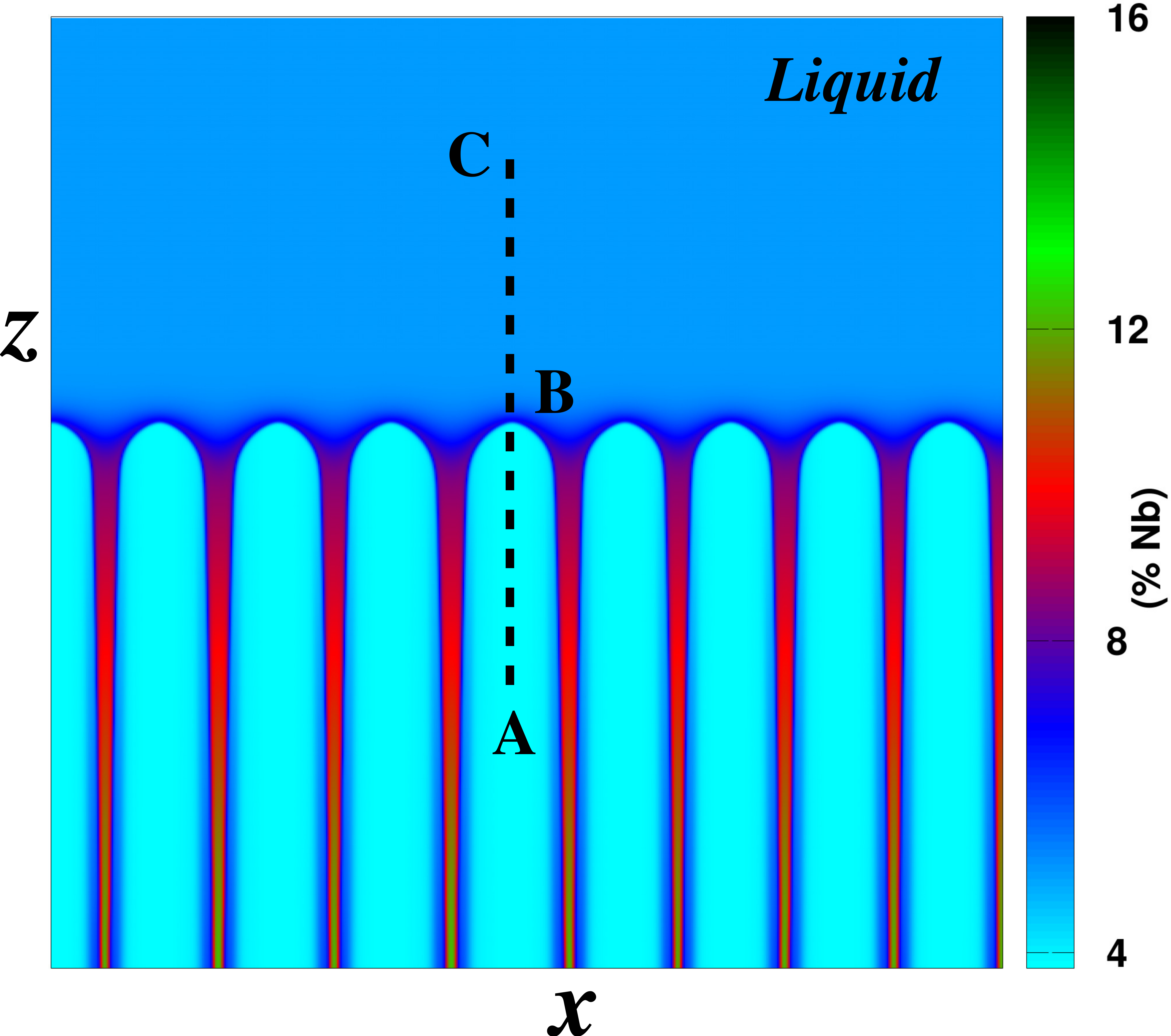}}\hspace{5mm}
\subfloat[]{\label{fig_profile}\includegraphics[scale=0.4]{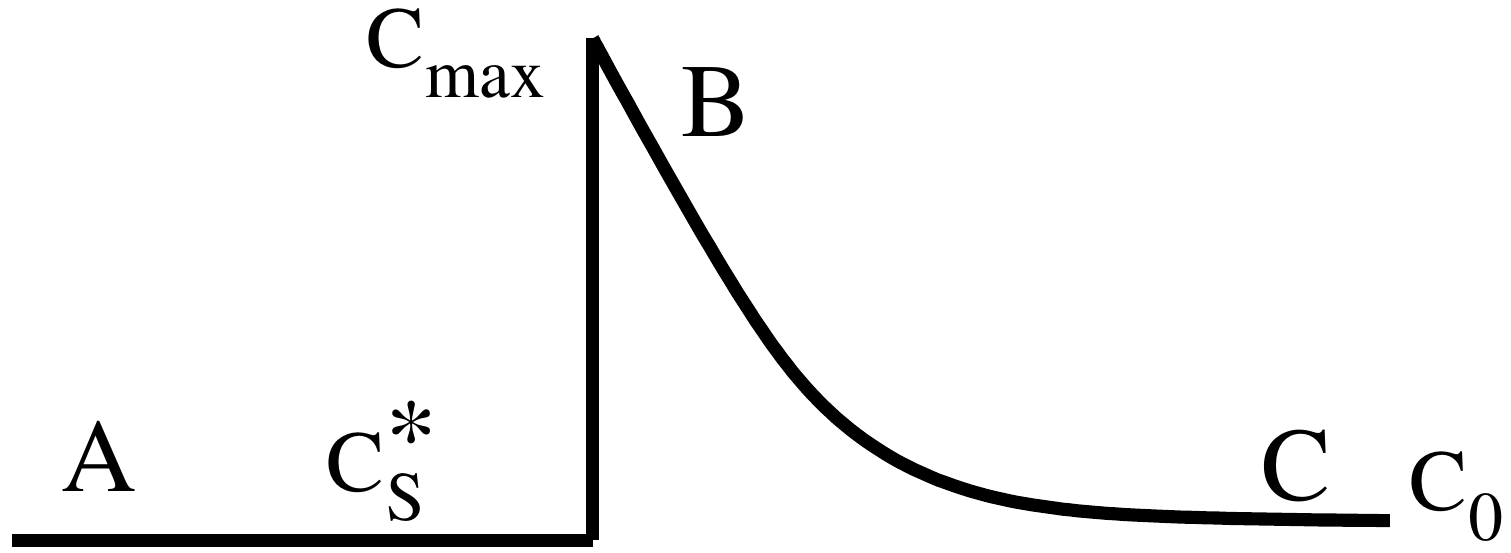}}
\caption{(a) Planar solid-liquid interface (not to scale) forms with a composition distribution that is shown using a color map. The corresponding Nb composition in both solid and liquid equals the nominal composition of the alloy $c_0$, and Nb enrichment at the interface is denoted by $c_{\text{max}}$. (b) Cellular interface that forms for a fixed value of ($G$, $V$) is illustrated using a color map. The characteristic compositions from A ($c_s^*$), B ($c_\text{max}$), and C ($c_0$) are determined. (c) Composition distribution across a plane front or a cell tip determines the scaled microsegregation number $k_v = c_s^{*}/c_\text{max}$. For a planar interface, $c_s^{*} = c_0$.}\label{fig_micro}
\end{figure}

The essence of the planar front and cellular solidification from the phase-field simulation is as follows. Simulation starts with the procedure described in Sec.~\ref{sec_parameters}. The initial perturbation in the solid-liquid interface either decays or grows (referred to as the Mullins-Sekerka instability~\cite{Mullins1964}) depending on the size of the domain and the strength of the perturbation, leading to either planar or cellular interface. Cellular solidification undergoes intermediate transient stages of growth by merging or splitting of the neighboring cells, which finally develop into a steady state cellular microstructure. At steady state, the solidification front grows with the temperature field at a constant velocity of $V$, which equals that estimated from the FEA simulation (refer Sec.~\ref{sec_parameters}). Typical simulations of planar and cellular interfaces are shown in Fig.~\ref{fig_micro}. 

Let us consider that $c_s^{*}$ is the composition of Nb in the solid and $c_\text{max}$ is the maximum composition of Nb in the liquid side of the interface. The scaled microsegregation is then determined by the solid composition ($c_s^{*}$) to the liquid composition ($c_\text{max}$) at the interface, given by $k_v = c_s^{*}/c_\text{max}$. For a planar interface, $c_s^{*} = c_0$ (nominal composition). For a cellular interface,  $c_s^{*} < c_0$ for a Ni-Nb alloy. Note that there is an uncertainty associated with the determination of $c_\text{max}$, this being a steeply peaked function, which we ignore in the present work. Since the diffusivity of the solute in the solid is several orders of magnitude smaller than that of the same in the liquid, we ignore diffusion in the solid and thus $c_s^{*}$ remains constant in the solid. In our simulations, the solid-liquid interface is not in local equilibrium, since the $k_v$ value extracted at the interface deviates from the equilibrium segregation coefficient $k$. In this context, an analytical approximation such as the Gulliver-Scheil equation~\cite{rappazbook} can predict the solute redistribution during planar solidification for the conditions of zero diffusion in the solid, infinite fast diffusion in the liquid, and local interface equilibrium. However, such an analytical approach would be even more strongly idealized compared to the present scenario where complex interface shapes form during non-equilibrium solidification and deviations from the phase-diagram are expected for arbitrary solidification conditions.

\subsection{Size effects}
\begin{figure}[h]
\centering
\subfloat[]{\label{fig_G}\includegraphics[scale=0.6]{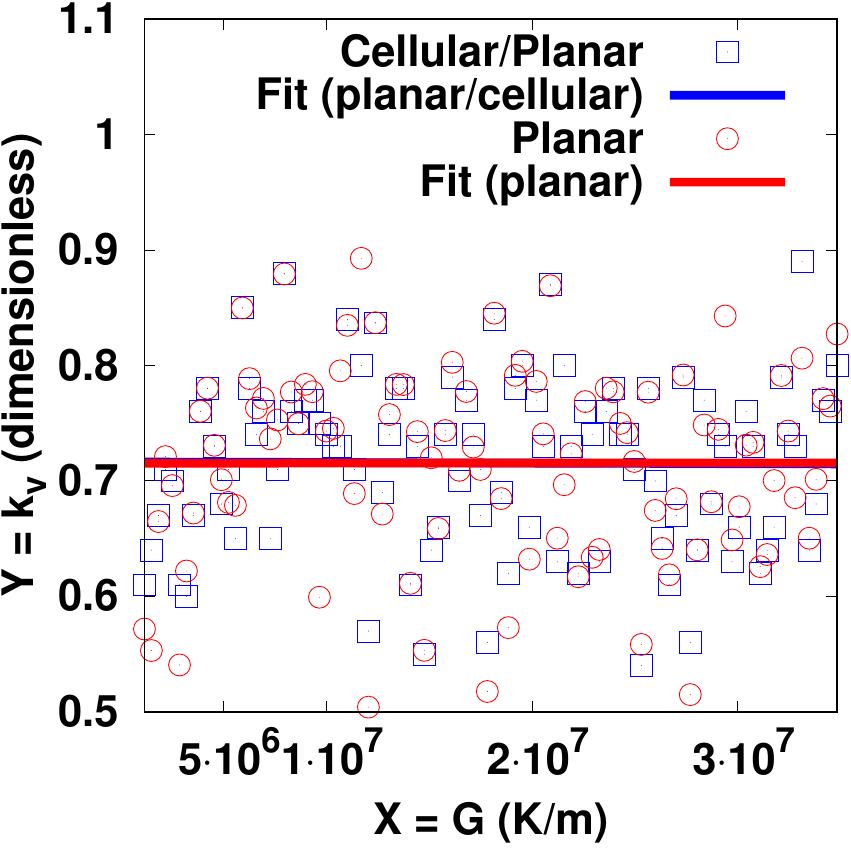}}\hspace{1cm}
\subfloat[]{\label{fig_V}\includegraphics[scale=0.6]{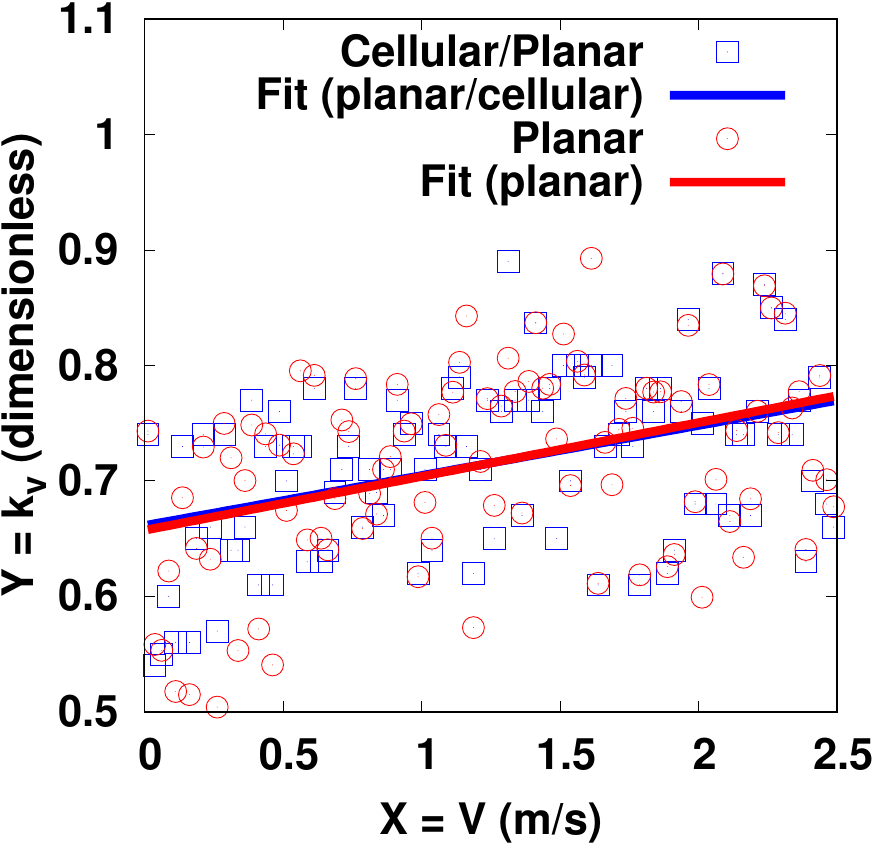}}
\caption{Scatterplots for (a) $G$ and $k_v$ (b) $V$ and $k_v$ for planar and cellular interfaces are presented. These phase-field data are extracted from two different domain sizes. Simulations with the small domain resulted into a planar interface, and the simulations with the larger domain resulted into either cellular or planar interface. The lines of best fit estimated from the simulation outputs from two independent runs overlap, signifying the same output distribution of microsegregation, on average.}
\label{fig_GVC}
\end{figure}

\begin{figure}[h]
\centering
\subfloat[]{\label{fig_D}\includegraphics[scale=0.6]{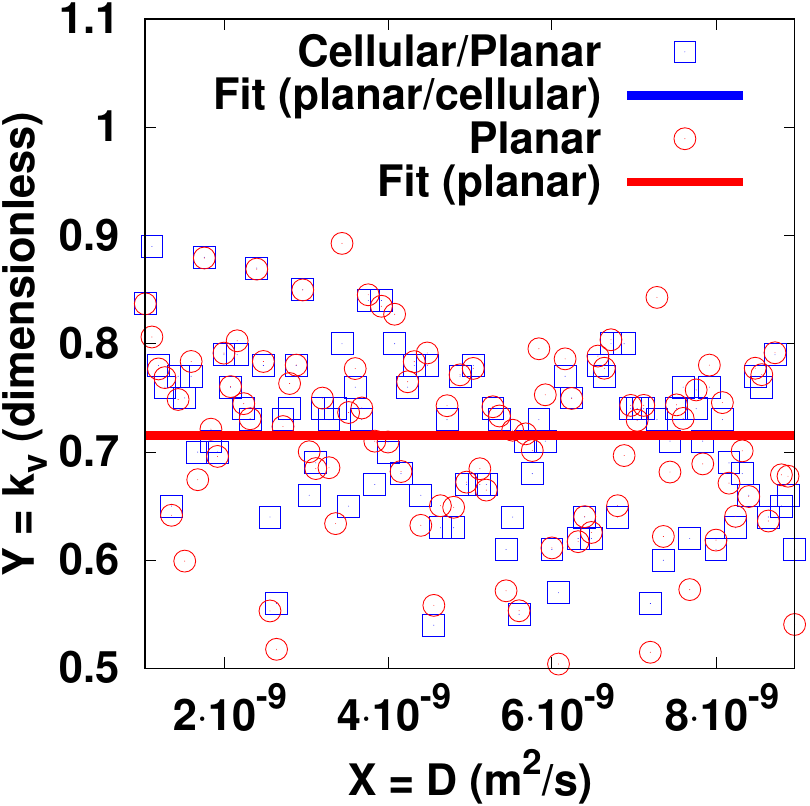}}\hfill
\subfloat[]{\label{fig_T}\includegraphics[scale=0.6]{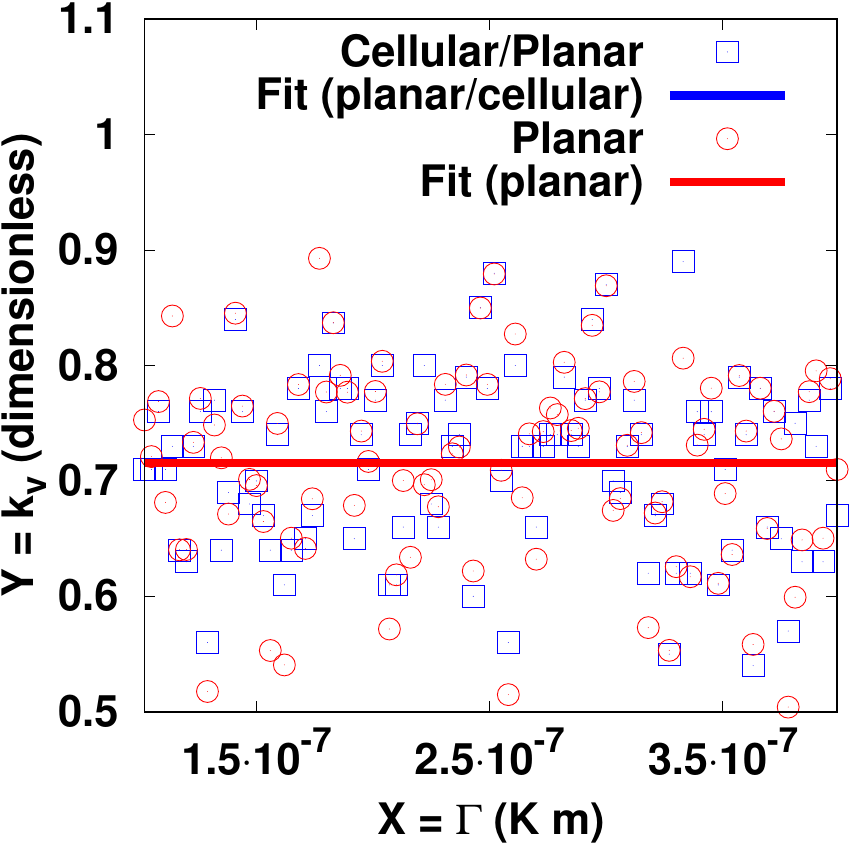}}\hfill
\subfloat[]{\label{fig_K}\includegraphics[scale=0.6]{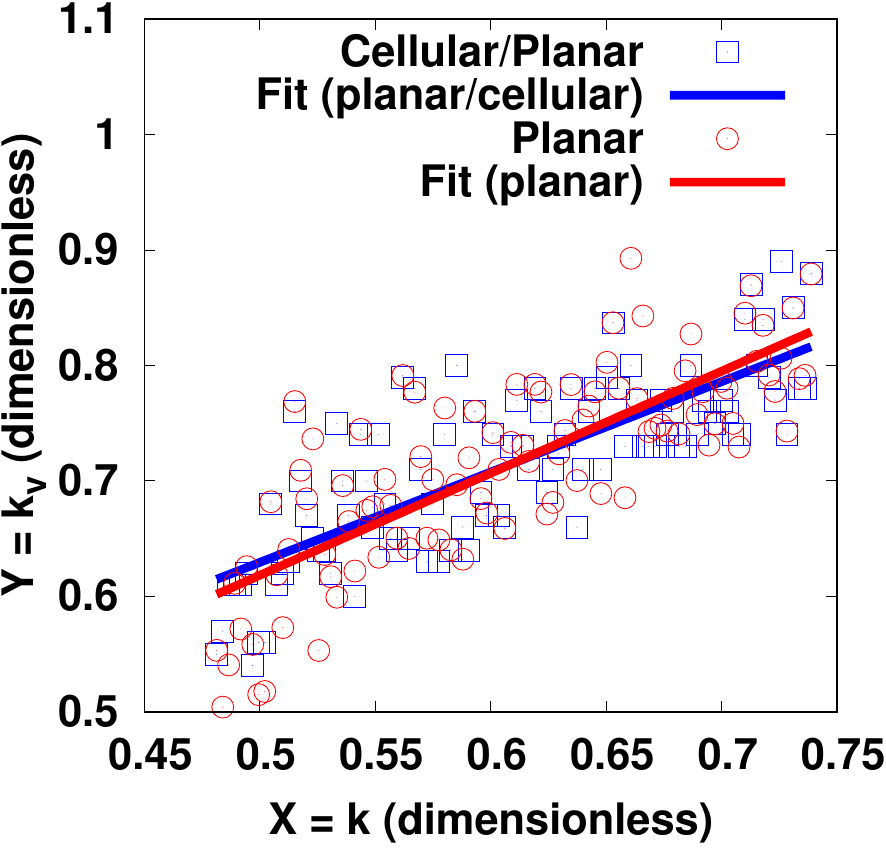}}
\caption{Scatterplots for (a) $D_l$ and $k_v$ (b) $\Gamma_{sl}$ and $k_v$ (c) $k$ and $k_v$ for planar and cellular interfaces are presented. These phase-field data are extracted from two different domain sizes. Simulations with the small domain resulted into a planar interface, and the simulations with the larger domain resulted into either cellular or planar interface. The lines of best fit estimated from the simulation outputs from two independent runs overlap, signifying the same output distribution of microsegregation, on average.}
\label{fig_DTK}
\end{figure}

\begin{figure}[h]
\centering
\subfloat[]{\includegraphics[scale=0.8]{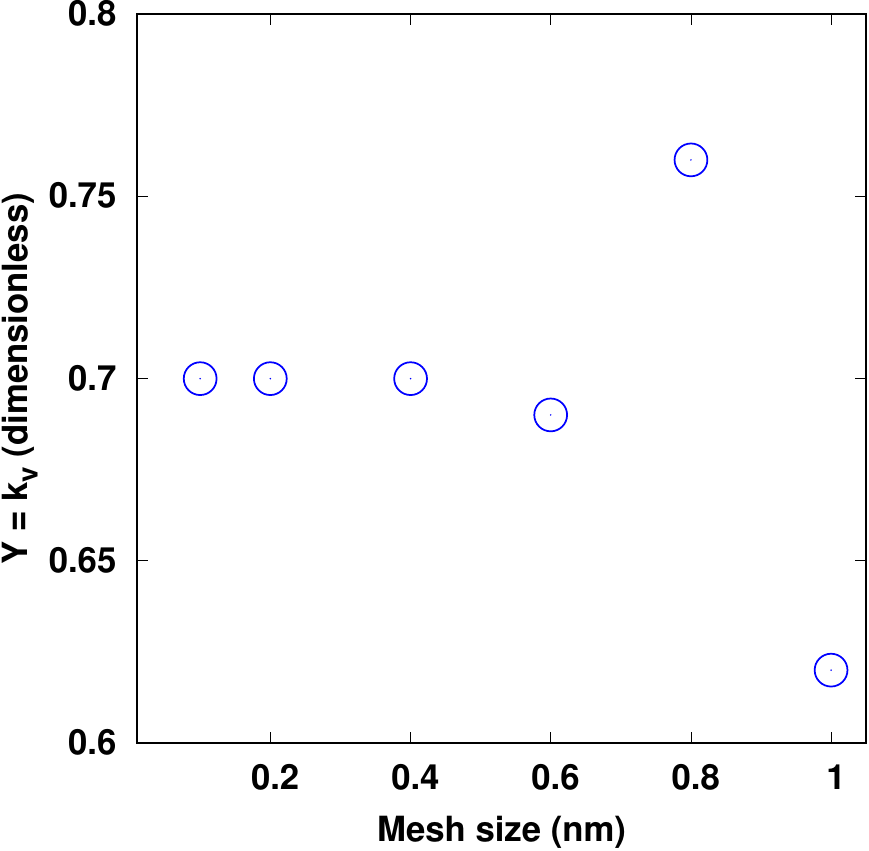}}\hspace{5mm}
\subfloat[]{\includegraphics[scale=0.8]{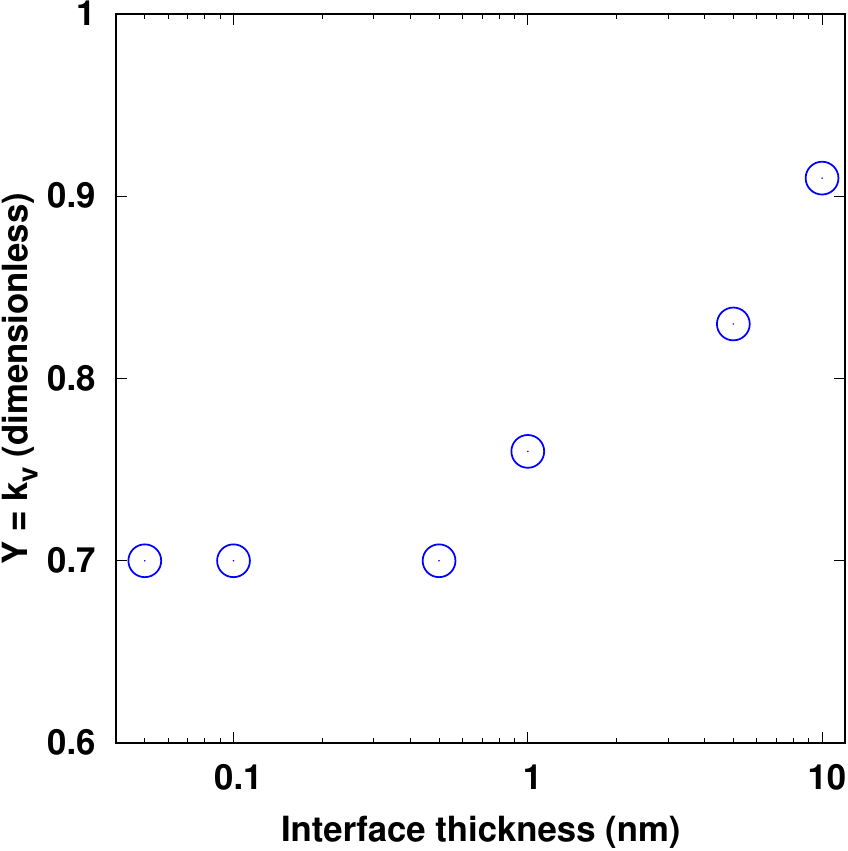}}
\caption{Phase-field simulations are performed for the maximum value of the velocity $V$ = 2.5 m s$^{-1}$ (for which maximum deviation from the interface equilibrium is expected) to illustrate the effect of the (a) mesh size and (b) interface width on Nb microsegregation. Simulated output $k_v$ values become grid-independent below a mesh size of 0.5 nm and interface thickness of 0.5 nm. Although not shown here, such a convergence study using the extreme values of other input parameters in Table~\ref{table_param} resulted in a similar observation.}\label{fig_size}
\end{figure}

To quantify the effects of the domain size on the simulated microstructures, we perform phase-field simulations for two lateral sizes ($N_x$) of the domain (refer to Sec.~\ref{sec_parameters}). The small domain (10 grid points) is used to obtain a planar solid-liquid interface in all simulations. Simulations with the large lateral domain (600 grid points) resulted into either planar or cellular interface depending on the combinations of $G$ and $V$ inputs. The simulated planar and cellular interfaces are presented in Fig.~\ref{fig_micro}. The values of $k_v$ are extracted from these interfaces and are plotted against the model inputs $\bm{X}= \lbrace\bm{x}_1,\ldots,\bm{x}_5 \rbrace = \lbrace G, V, D, \Gamma, k \rbrace$ in Figs.~\ref{fig_GVC} and \ref{fig_DTK}. It is clear that the $k_v$ extracted from planar and cellular solidification fronts are different when a single numerical experiment is considered that is performed with the same process and material parameters, but with different domain sizes. However, when the $k_v$ data is considered from the entire population of phase-field experiments, the $k_v$ distribution for both planar and cellular interfaces appears similar, on average, and the `average' lines that represent these two $k_v$ data set almost overlap with each other. Thus, the model output becomes size independent when the effects of the total population are considered. This is quantified using a sensitivity analysis in the following section. Since the distribution of $k_v$ population derived from both interfaces is similar, at least for the present ranges of the solidification conditions, henceforth, we present the output data from planar solidification only. For the sake of completeness, we present the effect of mesh size and interface width on the simulated $k_v$ values in Fig.~\ref{fig_size}. We find that below a mesh size of 0.5 nm and interface thickness of 0.5 nm simulation results become grid independent and therefore these numerical values are used for the present study.


\subsection{Sensitivity analysis}
For a sensitivity analysis, the scatterplots representing $k_v$ \textit{vs.} $\bm{X}= \lbrace G, V, D, \Gamma, k \rbrace$ are considered (Figs.~\ref{fig_GVC} and \ref{fig_DTK}). The effects of LPBF solidification conditions $\bm{X}^s = \lbrace G, V\rbrace$ on $k_v$ are shown in Fig.~\ref{fig_GVC}. The lines of best fit of the $k_v$ data extracted from both planar and cellular interfaces are presented. On average, $k_v$ varies between 0.5 and 0.9, which defines the microsegregation range of a Ni-Nb alloy during LPBF solidification. The $k_v$ data are plotted against the material parameters $\bm{X}^m = \lbrace D, \Gamma, k \rbrace$ in Fig.~\ref{fig_DTK}. Note that the line of best fit of the $k_v$ distribution remains `flat' for the changes in $\bm{X}^m = \Gamma$ (Fig.~\ref{fig_T}) and, also for, $\bm{X}^s = G$ (Fig.~\ref{fig_G}), meaning that there is no discernible pattern between each of these model inputs and the output. This is quantified using the correlation coefficient $\rho_{\bm{X}\bm{Y}}$.

\begin{figure}[h]
\begin{center}
\includegraphics[scale=0.9]{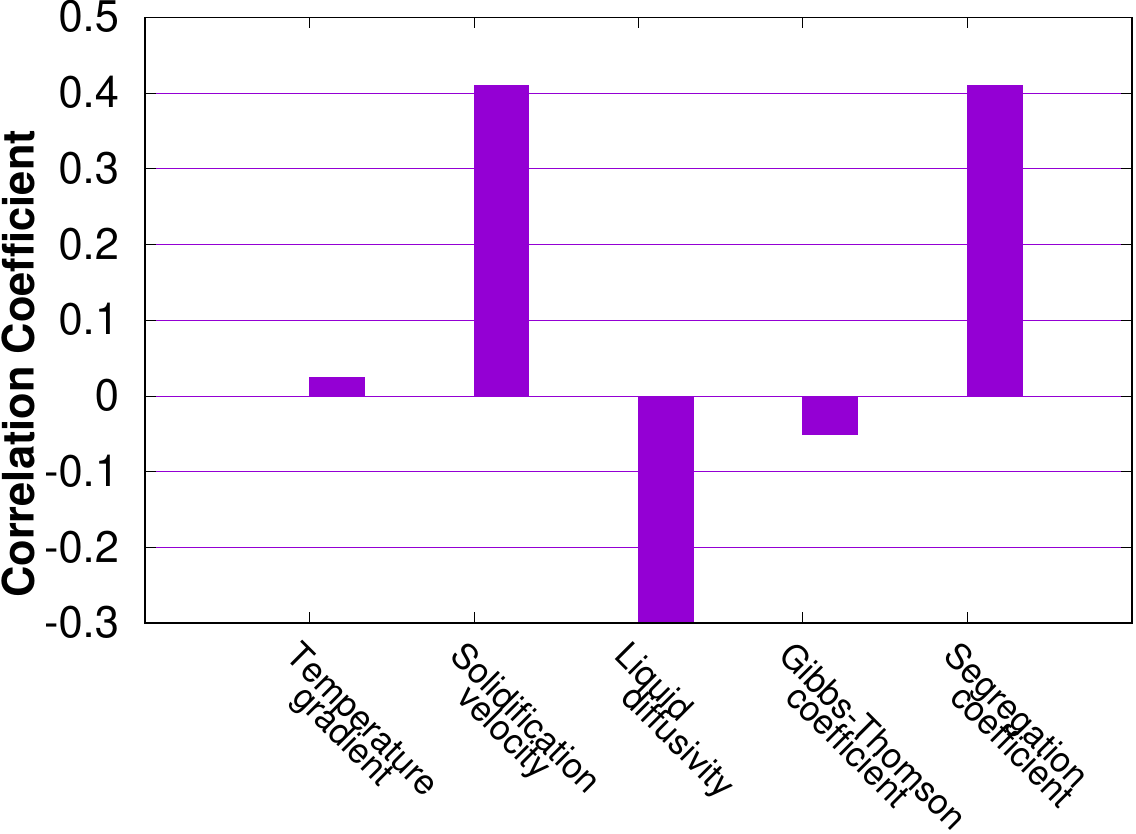}
\caption{A bar chart shows the calculated values of the correlation coefficient between various sources of uncertainty and the output. The solidification velocity, liquid diffusivity, and segregation coefficient are strongly correlated with the output, while temperature gradient and Gibbs-Thomson coefficient are not.}\label{fig_correlation}
\end{center}
\end{figure}

$\rho_{\bm{X}\bm{Y}}$ defines the direction and strength in a relationship between each input and the output. In order to quantify the sources of input uncertainty on $k_v$, we estimate $\rho_{\bm{X}\bm{Y}}$ following the procedure described in Sec.~\ref{sec_analysis}. The calculated $\rho_{\bm{X}\bm{Y}}$ value for each source of uncertainty is shown using a bar chart in Fig.~\ref{fig_correlation}. Note that the values of $\rho_{\bm{X}\bm{Y}}$ for $G$ and $\Gamma$ are close to $0$, which signify a negligible correlation between $k_v$ and $\bm{X}= \lbrace G, \Gamma \rbrace$ and explain the `flat' distributions in Figs.~\ref{fig_G} and \ref{fig_T}. Thus $k_v$ is not related in a meaningful way to the scores of $G$ and $\Gamma$ and it would be virtually impossible to predict $k_v$ simply by knowing the values of $G$ and $\Gamma$. On the other hand, the relationships between $k_v$ and ($\bm{X}^s= V$ and $\bm{X}^m = \lbrace D, k \rbrace$) are found to be correlated and are considered to be significant. The correlation between $k_v$ and $\bm{X}= \lbrace V, k \rbrace$ is positive, meaning that $k_v$ increases for the increase in $\bm{X}$ and \textit{vice versa}. In contrast, $D$ exhibits a negative correlation with $k_v$, meaning that $k_v$ increases when $D$ decreases and \textit{vice versa}. Niobium partitioning at the solid-liquid interface increases with the increasing rate of diffusion, decreasing $k_v$.

A liner regression analysis is used to predict the phase-field output for any given value of $\bm{X}$. Following the procedure described in Sec.~\ref{sec_analysis} and the calculation of $\rho_{\bm{X}\bm{Y}}$, it is now possible to obtain the regression equations (the units of $\bm{X}$ are in SI metric) that are expressed as:
\begin{eqnarray}\label{eq_regression} 
k_v &=& 0.047 \, V + 0.6625, \nonumber \\
k_v &=& -10^{7} \, D_l + 0.77, \, \text{and} \nonumber \\
k_v &=& 0.78 \, k + 0.24.
\end{eqnarray} 
These equations represent the lines of least square which are essentially the `fitted' lines in the scatterplots in Figs.~\ref{fig_GVC} and \ref{fig_DTK}. The slopes and intercepts of the regression lines are given in Eqs.~(\ref{eq_regression}). Note the nearly `flat' lines in Figs.~\ref{fig_G} and \ref{fig_T}, which signify that the $k_v$ data points are strongly scattered and bear no correlation with the input $\bm{X}$. Whereas in Fig.~\ref{fig_K}, $k_v$ is aligned with a slope, meaning $k_v$ is less scattered and a strong correlation between $k$ and $k_v$ exists.

\subsection{Surrogate analysis}\label{sec_surrogate}
The above correlation and regression analyses qualitatively predict the first order relationship between $\bm{X}$ and $\bm{Y}$. For a quantitative analysis of higher order relationships, brute force Monte Carlo simulations may be performed. However, the Monte Carlo method is computationally intensive for most computer simulation models. A surrogate model can be used instead for a quantitative assessment of the uncertainty in phase-field model inputs. Surrogate models often consist of Gaussian process or polynomial chaos emulators that provide computationally inexpensive approximations to the original computer model~\cite{allaire2010,ohagan2013}. We adopted a Gaussian process (GP) surrogate model in this work to substitute for the phase-field model and employed the framework described in Ref.~\cite{mahmoudi2018}, as implemented in Ref.~\cite{Mahmoudi2018MVCalibration}. A brief outline of the GP model and the procedure required to obtain the surrogate predictions are presented in~\ref{appendix_surrogate}. 

\begin{figure}[h]
\centering
\subfloat[]{\label{fig_surrogate3}\includegraphics[scale=0.97]{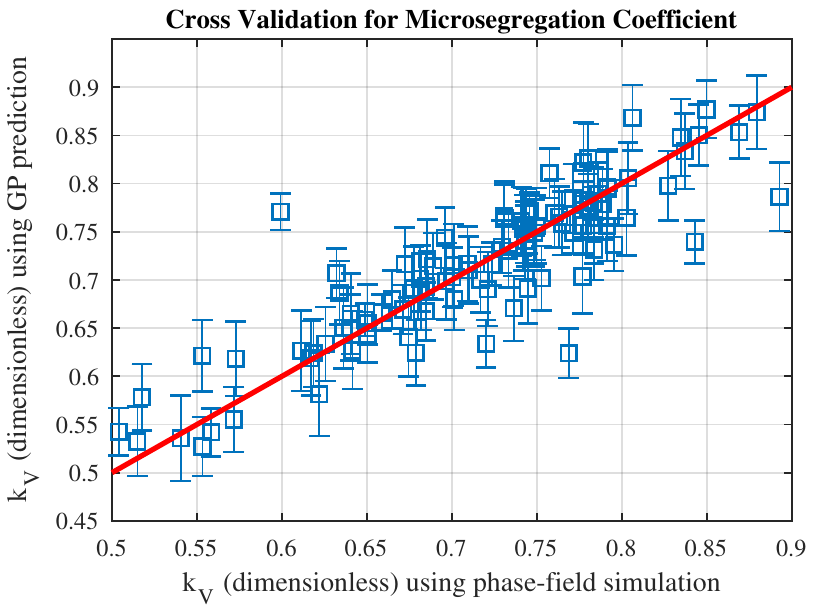}}\hspace{1cm}
\subfloat[]{\label{fig_error}\includegraphics[scale=0.7]{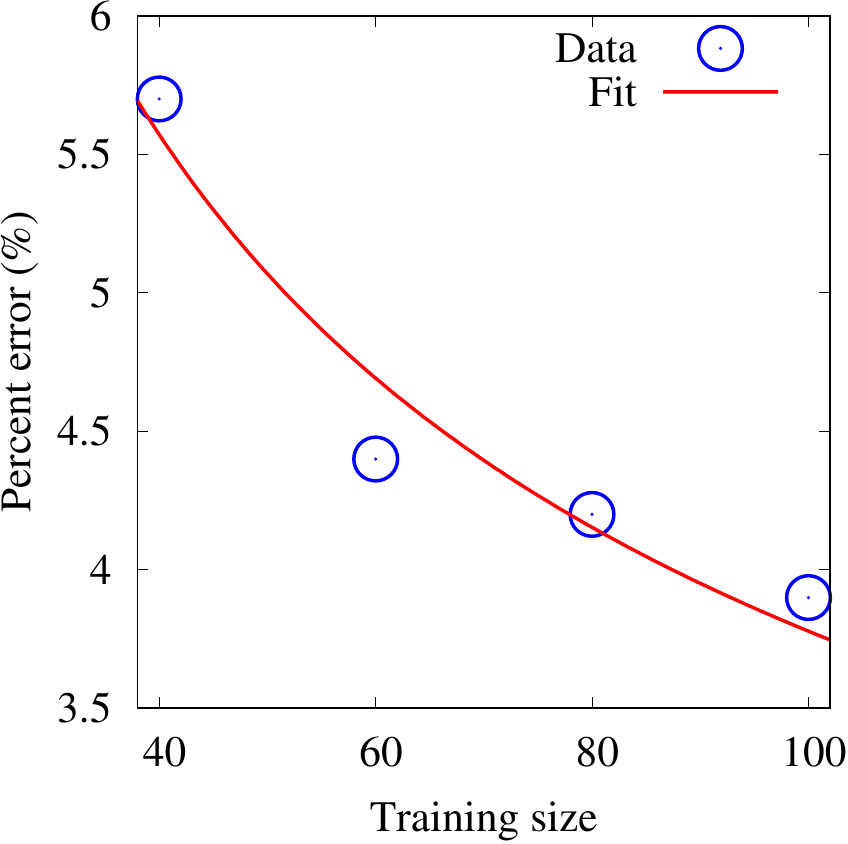}}
\caption{(a) A Gaussian process surrogate model predicts microsegregation using 3 uncertain inputs $\bm{X}= [V, D, k]$. The horizontal axis represents the phase-field model predictions and the vertical axis represents the surrogate model predictions. Red line represents the ideal case when surrogate model predictions are in full agreement with the phase-field predictions. Each surrogate prediction is plotted with a confidence interval that represents the standard deviation around the mean. On average, the difference between the surrogate model and the phase-field predictions is satisfactory. (b) Percentage error between these predictions is shown as a function of the training data size. As expected, the difference between the training data (i.e., phase-field) and surrogate prediction decreases with increasing data size.}
\label{fig_surrogate}
\end{figure}

The surrogate model is constructed to predict the output $k_v$ for a given input $\bm{X}$. Two different GP surrogate models run $N$ computational experiments with varying values of two different input sets, one with 5 variable inputs $\bm{X}= \lbrace G, V, D, \Gamma, k \rbrace$ and the other with 3 inputs $\bm{X}=  \lbrace V, D, k \rbrace$, the parameters of which were considered significant after the correlation analysis. The corresponding surrogate model predictions and their confidence intervals are presented in Fig.~\ref{fig_surrogate3}, along with the ideal case (red line) that represents the phase-field predictions. We do not present the results of the 5-parameters surrogate model, which shows a similar behavior as in Fig.~\ref{fig_surrogate3}; the statistical quantities estimated using both models are given in Table~\ref{table_mape}. For a quantitative assessment, the computed mean absolute predictive (Eq.~(\ref{eq_mape})) and percentage errors for both surrogate models are reported, the statistical measures of which indicate satisfactory performance of the surrogate models used. Note that the error metrics determined for the surrogate model with 3 inputs and the surrogate model with 5 inputs are similar, signifying that the effects of variability regarding the two additional inputs in the 5-parameters model are negligible, on average. The GP surrogate model with 5 inputs has a slightly lower performance compared to the model with 3 inputs, which may be due to the over-fitting caused by more inputs compared to the model with 3 inputs. These results signify that the three parameters, $V$, $D$, and $k$, represent the primary sources of uncertainty in the simulation of LPBF solidification process, whereas the variability in two inputs, $G$ and $\Gamma$, can be neglected safely in order to reduce computational effort required for quantification. 

The mean absolute percent difference between the phase-field and surrogate model outputs is 3.9\% (Table.~\ref{table_mape}). One of the primary reasons for this discrepancy can be due to a small number ($N = 100$) of samples used to train the surrogate model. Figure~\ref{fig_error} shows the effect of the training data size on the percent error, which can be expressed using an empirical power law of percent error = 0.266 $\times$ size$^{-0.424}$. In our opinion, $N = 100$ represents the lower limit of the number of samples required to apply uncertainty analysis in materials research. More training samples are needed for a more accurate prediction of the surrogate model used in the present study.
\begin{table}[h]
\centering{}%
\begin{tabular}{lcc}
\hline \hline
 Statistics & GP with 3 inputs & GP with 5 inputs  \\
\hline
Mean absolute predictive error (MAPE) & 0.0273 &  0.0287 \\
Observed range in simulation & 0.39 & 0.39 \\
Mean absolute percent error & 3.9\% & 4.1\%  \\
\hline \hline
\end{tabular}
\caption{Mean absolute predictive and percent errors of the surrogate model predictions.}\label{table_mape}
\end{table}%

\subsection{Frequency analysis}
\begin{figure}[h]
\centering
\subfloat[]{\label{fig_cdf}\includegraphics[scale=0.6]{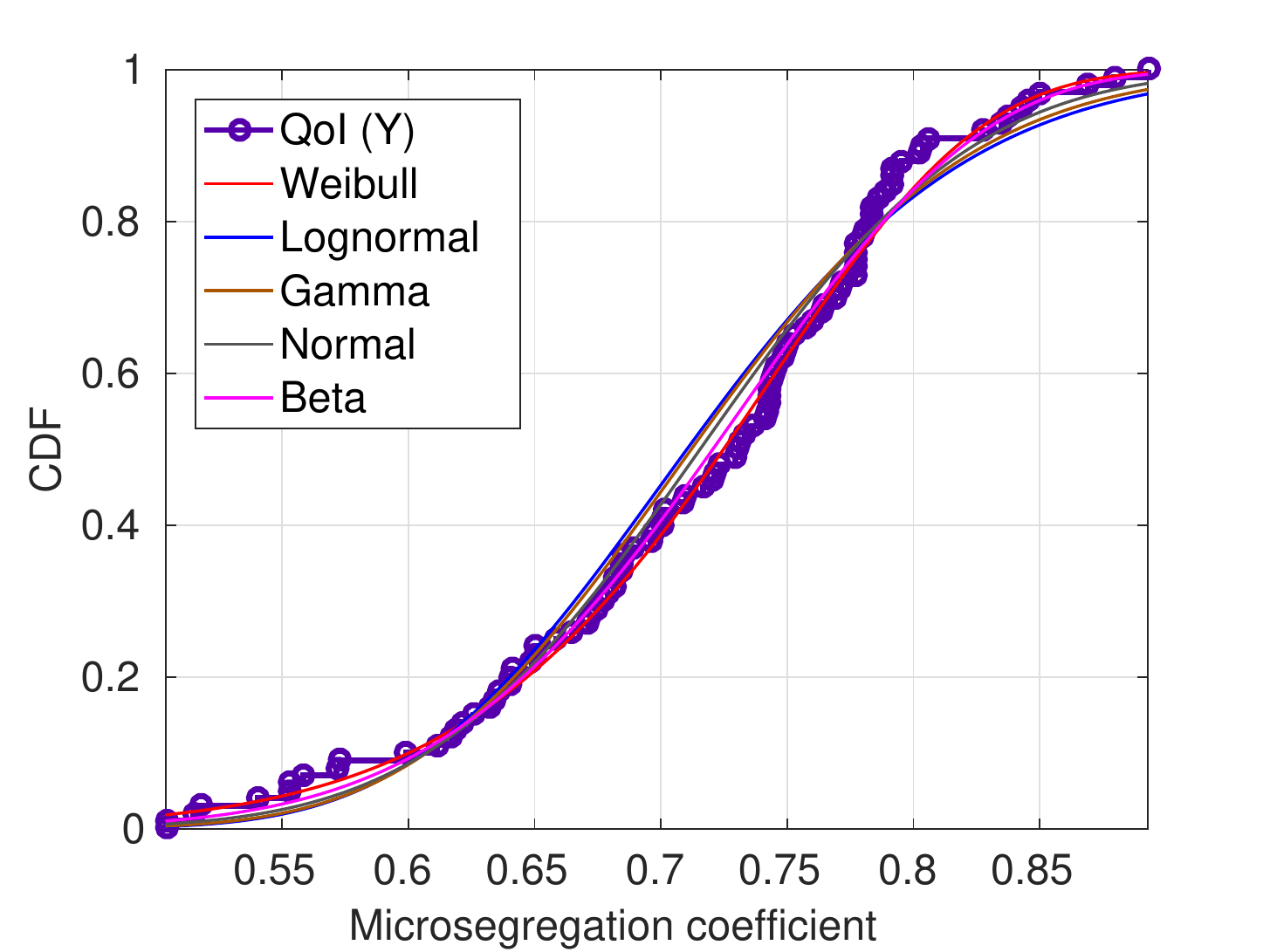}}\hfill
\subfloat[]{\label{fig_mle}\includegraphics[scale=0.6]{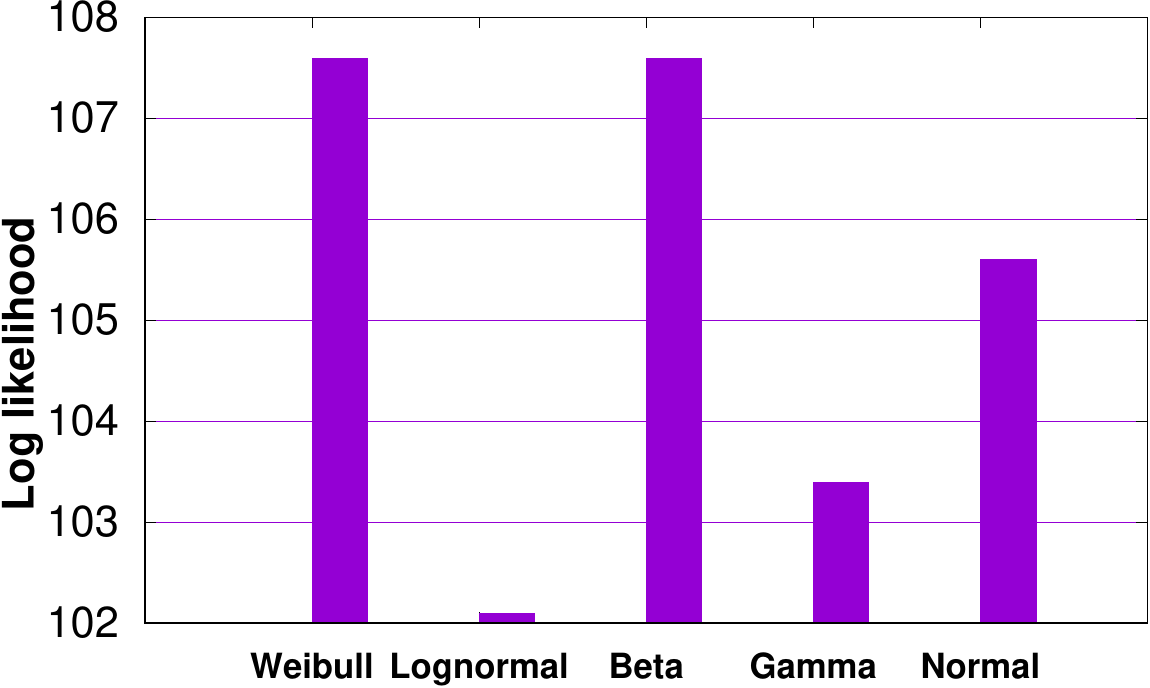}}
\caption{(a) Calculated cumulative distribution of the phase-field simulated $k_v$ population ($N=100$) is fitted against different target CDFs. (b) Maximum `log-likelihood' of each fitted CDF against the empirical CDF is shown.}
\label{fig_cdf_all}
\end{figure}

\begin{figure}[h]
\begin{center}
\includegraphics[scale=0.8]{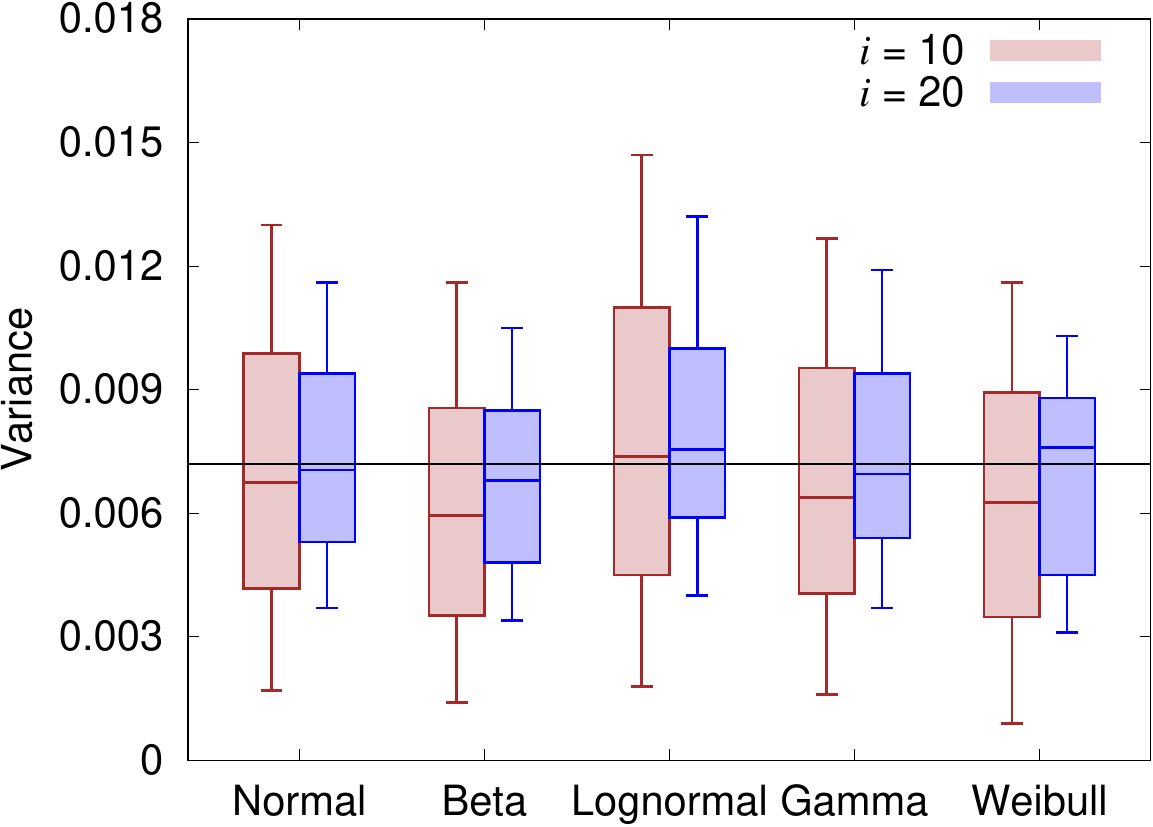}
\caption{Phase-field simulated microsegregation distribution is fitted against different target CDFs for two samples of size $i = 10$ (left bar) and $i=20$ (right bar).}\label{fig_ncdf}
\end{center}
\end{figure}

We used cumulative distributions and probability densities to analyze the $k_v$ distribution obtained from our phase-field simulations. Following the procedure described in Sec.~\ref{sec_analysis}, $k_v$ measurements are ranked in the ascending order, as presented in Table~\ref{table_cdf}, where $F$ is the cumulative distribution. The corresponding empirical CDF is presented in Fig.~\ref{fig_cdf}. To gain statistical insights from the $k_v$ data, we have fitted our empirical CDF with several theoretical CDFs that include normal, log-normal, gamma, beta, and Weibull functions. Weibull and beta functions produced maximum `log likelihood' with the empirical $k_v$ (Fig.~\ref{fig_mle}). Further, to test the effects of sampling, we have used these target CDFs to fit two different random samples of $k_v$ of size $i=10$ and $i=20$. We use an objective function: $\frac{1}{N}$ $\sum_i^{N} [F_i - \hat{F}_{i}^{t}]^2$, where $F_i$ is the empirical CDF value and $\hat{F}_{i}^{t}$ is the corresponding fitted CDF value of a distribution $t$, to measure how good the average `fit' is between the two. The corresponding statistical measurements are presented in Fig.~\ref{fig_ncdf}. Clearly, the median of the variance for the log-normal and normal distributions is closest to the $k_v$ distribution for $i=10$ and $i=20$ samples, respectively. However, the spread and the range of the $k_v$ distribution are closest to a beta CDF, on average.
\begin{table}[h]
\begin{center}
\begin{tabular}{l c c}
\hline \hline
Rank ($m$) & Output ($k_v$) & Relative Rank ($F$) \\
\hline
1 & 0.504062 & 0.01 \\
2 & 0.514948 & 0.02 \\
3 & 0.517629 & 0.03 \\
4 & 0.540556 & 0.04 \\
5 & 0.553158 & 0.05 \\
..\\
95 & 0.842911 & 0.95 \\
96 & 0.845357 & 0.96 \\
97 & 0.849884 & 0.97 \\
98 & 0.869146 & 0.98 \\
99 & 0.879405 & 0.99 \\
N = 100 & 0.892838 & 1.0 \\
\hline \hline
\end{tabular}
\caption{Ranked ordered phase-field measurements.}\label{table_cdf}
\end{center}
\end{table}%

\begin{figure}[h]
\begin{center}
\includegraphics[scale=0.75]{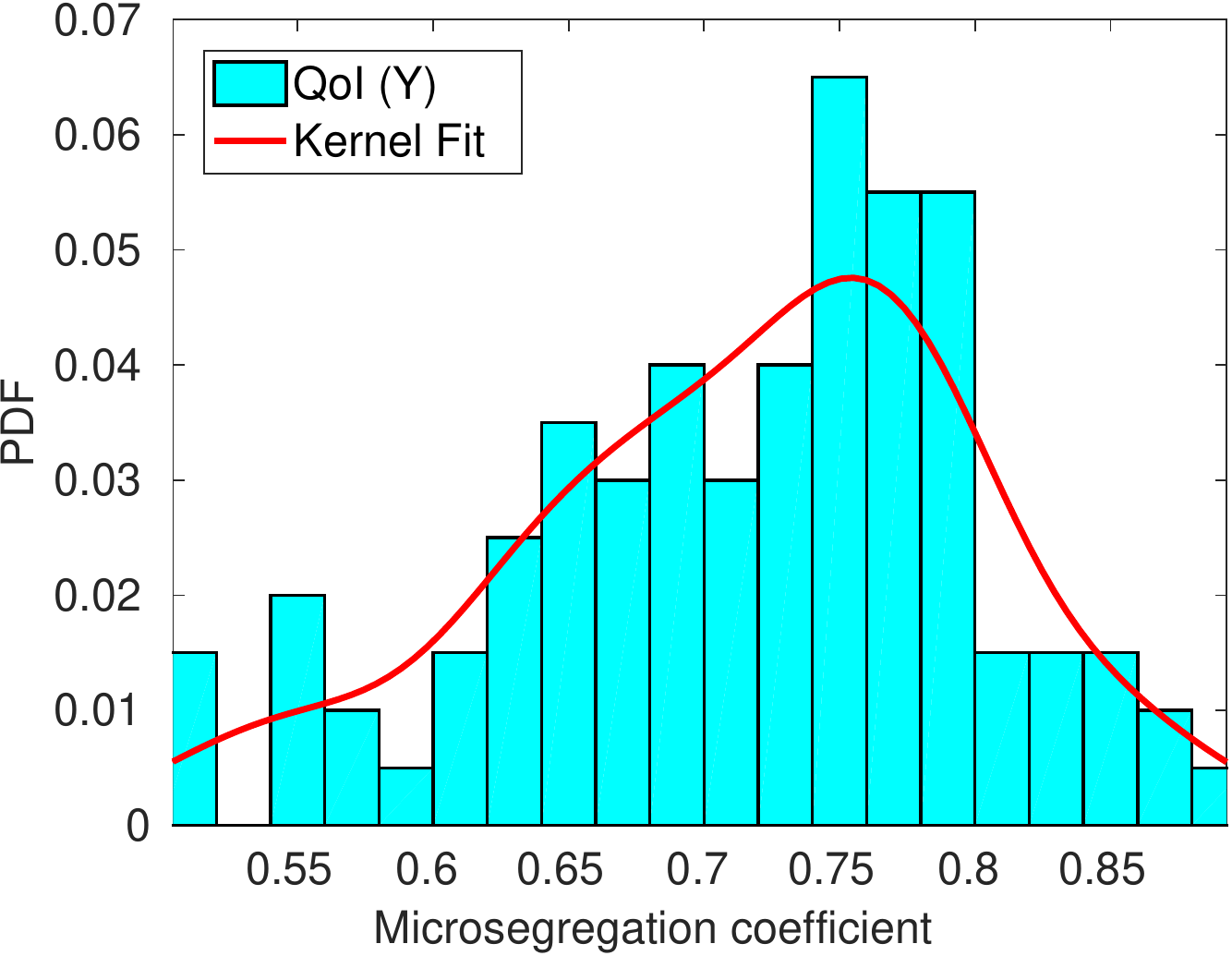}
\caption{Histogram of the microsegregation data and the fitted PDF using a normal kernel is illustrated.}\label{fig_pdf}
\end{center}
\end{figure}

Figure~\ref{fig_pdf} plots the histogram of the $k_v$ data and the associated fitted PDF using a normal kernel. The peak of the histogram is close to the nominal $k_v$ that represents the characteristics microsegregation value of an as-built Ni-Nb alloy during the LPBF process. The significance of such a PDF is twofold. The distribution can be compared with experimental measurements for a calibration of the model parameters in an inverse UQ approach (Fig.~\ref{fig_schematic}). Also, it can be used to model the subsequent heat treatment processes to aid the microstructural design in a forward UQ problem.

\section{Discussion}\label{sec_discussion}
\subsection{General remarks}
We present a case study illustrating how different sources of model uncertainty at the mesoscale can be identified and subsequent frequency analysis can be performed to understand the microsegregation statistics that result during the microstructure evolution in the LPBF process. We used an alloy phase-field model~\cite{Echebarria2004} as a reference to simulate the solidification behavior of a dilute Ni-Nb alloy (far away from Ni-Nb eutectic composition). The model explicitly tracks the solidification front, from where the Nb composition was extracted to determine the microsegregation number $k_v$. Note that our phase-field model is based on significant simplifications of the LPBF process and ignores the effects of several melt pool physics, such as the diffusion of heat and the convection in the liquid, on $k_v$. A justification for the use of these approximations to keep our phase-field simulations tractable is discussed in Ref.~\cite{supriyo20173d}. Ours is a baseline reference study to understand the variability of the model inputs on the microstructural features in LPBF solidification regime. Although there are process parameter ($G$, $V$) uncertainties that originate during FEA simulations, we do not perform a multi-model-level UQ~\cite{mahadevan2012} in this work and focus only the influence of phase-field single-model-level UQ on microstructure statistics.

The computational requirements of the phase-field model depend on the approximations used and the domain size employed to make the simulations tractable. For example, simulating multicomponent solidification with the incorporation of additional physics such as fluid flow cost significantly more computational resources compared to simulating binary alloy solidification with reasonable approximations. Although the phase-field equations of motion (Eqs.~(\ref{eq_phi}) and (\ref{eq_c})) scale well on large parallel computer architectures, these simulations often run for days on a multi-node multi-core supercomputing framework to reach a steady state with a statistically appreciable number of microstructural features that further depend on the domain size used in the simulations.

Validation of our phase-field model against both analytical solutions and, where available, experimental measurements in the LPBF solidification limit have been reported previously~\cite{Trevor2017,supriyo20173d} and is therefore not repeated here. In these previous works, the concentration variation of Nb in the solidified melt pool could not be resolved during a scanning electron microscopy analysis since the beam spot size was quite large with respect to the extremely fine microsegregation features. Niobium concentration at the solid-liquid interface varies roughly between 5 wt\% and 12 wt\% in our simulations (Fig.~\ref{fig_micro}). As far as we are aware, the most closely related experimental measurement of elemental distribution during LPBF is reported in Figure~9(a) within Ref.~\cite{ranadip2017}. These experiments however correspond to a fixed $G$ and $V$, for which Nb varied between 5 wt\% and 10 wt\% along the traces of the solid-liquid interface in a solidified IN718 molten pool. When its binary analog is considered, as in the present case, Nb partitioning across the interface can be more pronounced since the interactions among other alloying elements \textit{via} solidification range, diffusivity, and partition coefficient are absent.

\subsection{Influence of parameters}
The majority of our phase-field simulations resulted in planar solidification, and the remaining simulations predicted cellular solidification. In the literature~\cite{ghosh2018single,Ghosh2018,vrancken2014,tao2019}, LPBF experiments were performed for a single set of laser parameter, for which the resulting solidification conditions yielded primarily cellular microstructure. The ranges of laser parameters that provided the ranges of solidification conditions in our phase-field simulations are large when compared to an experimental study that uses a single set of laser parameter. As a result, the resultant ranges of $G$ and $V$ were large enough to produce combinations of cellular and planar interfaces. The simulated morphologies are consistent with the conditions of $V_{cs}$ and $V_{ab}$ (discussed in Sec.~\ref{sec_general}); since the majority of the $V$ samples are higher than $V_{ab}$, the corresponding interface morphology is planar. Such broad distributions of laser parameters and the resultant solidification conditions, on average, approximate the LPBF process and solidification maps of a generic alloy. Our results will be valuable when compared directly with uncertainties in cellular and dendritic morphologies that result for the LPBF solidification conditions. Work in these directions is currently in progress.

$G$ and $V$ are estimated together at a particular location in the melt pool and are therefore correlated regarding the location-specific microstructural requirements during solidification. Single-track or multiple-track laser melting processes affect the solidification conditions as the neighboring melt pools interact. In reality, there is also a certain probability of finding the same value of either/both $G$ and $V$ at several locations in the melt pool, since these conditions primarily vary as a function of the melt pool height in a single molten pool within multiple melt pools (Fig.~\ref{fig_process}c). As a consequence, when these melt pools solidify into a microstructure, the key microstructural features may remain similar across microscopic distances. Similarly, when multiple melt pool microstructures are considered in that both of the solidification conditions may be repeated across these microstructures. Further, our analysis proves that in the high-velocity limit (or, at least where the majority of our simulation condition falls) $G$ has only a minor contribution to the time-dependent solidification process. In this limit, on average, independent distributions of $G$ and $V$ can effectively assess the microstructure space that forms during LPBF. For a more accurate description of solidification, however, $G$ and $V$ may be shuffled together with a correlation during Latin Hypercube sampling.

Epistemic model solution uncertainties include numerical parameters such as interface width and mesh size. Convergence studies were performed in order to use appropriate values of interface width of 0.5 nm and mesh size of 0.3 nm, for which the simulated microstructures became grid independent. This way, numerical uncertainties were kept to a minimum. Model input uncertainty sources can have significant contributions to output uncertainties, and the inputs to which model and/or microstructure are sensitive are $V$, $D$, and $k$. These are the aleatoric sources present in the modeling framework considered here. The temperature gradient $G$ and the Gibbs-Thomson coefficient $\Gamma$ do not relate to $k_v$ in any meaningful way, and hence they can be fixed at certain specific values, e.g., mean, without considering their variability in practice (epistemic uncertainties).

The contribution of solidification conditions on the microstructure evolution in LPBF solidification regime can be ranked after the measurements of Pearson correlation. This analysis shows that $V$ ($\rho_{\bm{X}\bm{Y}} = 0.42$) is the most important process parameter and the effects of $G$ are negligible ($\rho_{\bm{X}\bm{Y}} =  0.03$). This explains why many researchers~\cite{Trevor2017,Boettinger1999} ignored the variability in $G$ and varied the value of $V$ during the modeling of solidification at the high-velocity LPBF limit. Note that, in this first approach, we have only determined the contribution of an individual input on the output QoI without considering the mutual interactions among the inputs. One can use the Sobol' variance analysis~\cite{li2016_sobol} to estimate such mutual statistical interactions among the inputs which is given by:
\begin{equation}
S_i^T = 1 - \frac{E_{\bm{X}_{\sim i}}\left(\text{Var}_{x_i}\left(k_v|{\bm{X}}_{\sim i}\right)\right)}{\text{Var}\,(k_v)},
\end{equation} 
where $x_i$ is the $i$-th input variable, ${\bm{X}}_{\sim i}$ is the vector of variables excluding $x_i$, $\text{Var}\,()$ is the variance, and $\text{Var}_{x_i}\left(k_v|\bm{X}_{\sim i}\right)$ is the variance by freezing $x_i$. Note that the Sobol' method would require an unaffordable amount of phase-field simulations that consume significant computational time and resources. Therefore, a suitable surrogate model such as Gaussian process model~\cite{mahadevan2017} and polynomial chaos expansion~\cite{tapia2018} may be used to approximate the expensive phase-field calculations. Following the correlation analysis, a regression analysis was used to represent the linear relationships between each input and the output. Regression equations give an intuitive experience by providing the formula for calculating the predicted value of the output when an actual value of the input variable is known. This is our first approach, and hence we ignored statistical interactions among the input variables that can be estimated using multiple regressions~\cite{aiken1991}.

\subsection{Surrogate modeling perspectives}
A Gaussian process (GP) surrogate model was used previously to replace the FEA simulations in Refs.~\cite{mahmoudi2018,mahadevan2018}. There has been no attempt yet to use a surrogate model to quantify the process-structure-property linkages that develop during an AM process~\cite{popova2017,jung2019}. Our approach to approximate the phase-field model using a Gaussian process regression model is the first step in that direction. The surrogate model predictions are satisfactory (at least for the given sample size) as they successfully replicate our phase-field predictions (Fig.~\ref{fig_surrogate}). A polynomial chaos expansion~\cite{tapia2018} may be used instead of GP as a surrogate base. However, a GP surrogate model is more efficient compared to the polynomial chaos expansion in terms of capturing local structures, flexibility, and quantifying uncertainty~\cite{ohagan2013}. 

Constructing our surrogate model was based on approximating a physics-based model using a multivariate GP model~\cite{mahmoudi2018}. To achieve this, a number of simulations from the computationally expensive physics-based model must be generated first that uses a sampling technique such as the Latin Hypercube Design. One direction in the literature to improve the prediction of the surrogate model is to construct it using a combination of high-fidelity and low-fidelity simulations, rather than only using high-fidelity simulations. For example, construction of a surrogate model that uses complex high-fidelity phase-field solidification model and lower-fidelity simulations from an analytical model may train the surrogate model faster compared to generating the training points only from a high-fidelity model. Even though low-fidelity simulations on their own do not provide good predictions like the original high-fidelity physics-based model, they can be very beneficial in obtaining better surrogate model predictions than the case that only bases the surrogate model on high-fidelity simulations~\cite{kennedy2000,tuo2014}. Further, to optimally improve the predictive capability of the GP surrogate model over the domain we could develop sequential full model sampling policies based on maximizing the Kullback-Liebler divergence (Eq.~(\ref{eq_kl})) between the current surrogate model and a surrogate model with one extra data point~\cite{huan2014gradient,ghoreishi2018multi}. In this sense, the process would maximize the information gained on every query to the expensive full model.

\subsection{Microsegregation perspectives}
Obtaining a quantitative basis of the microsegregation during LPBF solidification is far from straightforward. In this regime, a finite level of microsegregation is expected that is confirmed by our phase-field simulations. The estimated levels of microsegregation, however, may depend on how the anti-trapping flux term (Eq.~(\ref{eq_c})) is implemented in the present model. Anti-trapping flux was introduced by Karma~\cite{Karma2001} using a thin interface analysis to eliminate the artifacts due to the use of large numerical interface width during the phase-field simulation for a `magic' value of $a_t = 1/(2\sqrt{2})$. A value of $a_t$ larger than this may lead to the measured $k_v$ to be lower than its equilibrium value $k$ and therefore becomes unphysical. To a good approximation, the results presented here can be considered close to the lower limit when the phase-field model in Refs.~\cite{Karma2001,Echebarria2004} was used as a reference to simulate the ranges and distribution of $k_v$ in the LPBF regime. A more systematic work is necessary, as analyzed in Refs.~\cite{mullis2010,ohno2009}, to explore how the inclusion of $a_t$ within our phase-field model affects the measured levels of $k_v$ for a more accurate, physical description of solute trapping.

\begin{figure}[h]
\centering
\includegraphics[scale=0.8]{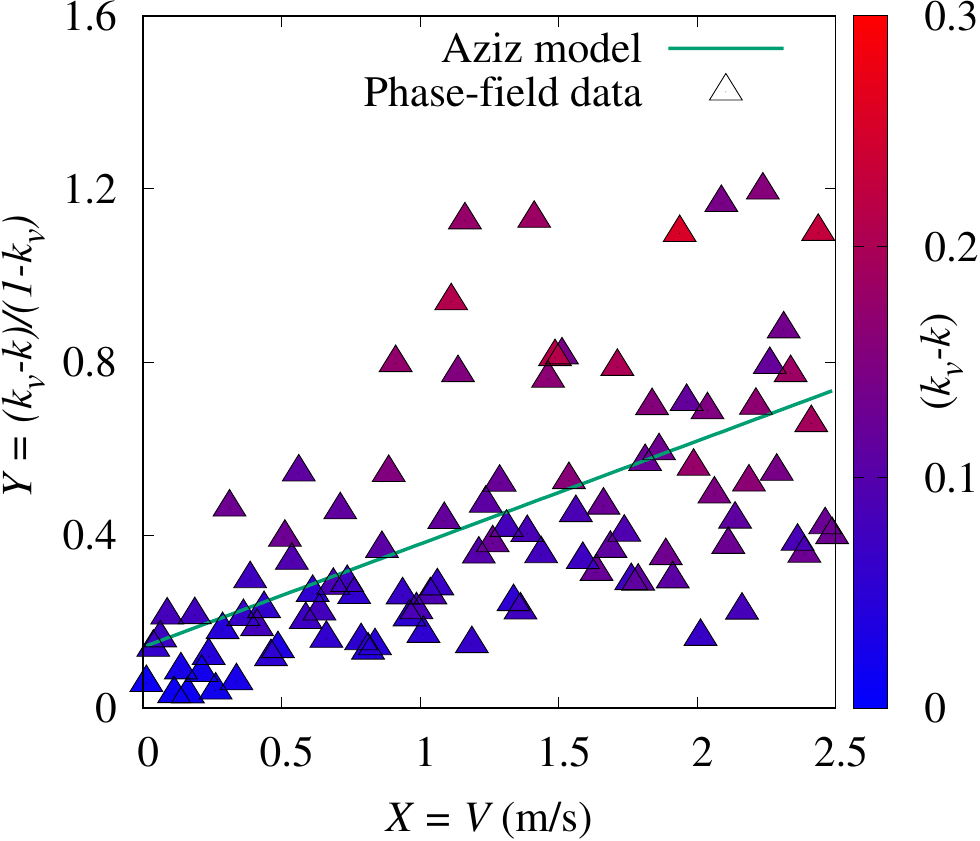}
\caption{The map of Nb partitioning at the solid-liquid interface during LPBF is illustrated as a function of $V$. On average, the magnitude of solute trapping or the deviation from the local equilibrium (proportional to $k_v - k$) increases with increasing $V$ that is superimposed on the phase-field data using a color map. A linear regression model is fit to the data following the Aziz model~\cite{Aziz1982}.}\label{fig_aziz}
\end{figure}

In experimental solidification studies, the microsegregation coefficient $k_v$ is used to determine the characteristic interface diffusion velocity $V_D = D/W$, where $D$ (Eq.~(\ref{eq_c})) is the diffusivity of the liquid and $W$ (Eq.~(\ref{eq_phi})) is the physical width of the solid-liquid interface. $V_D$ determines the solute trapping behavior of an alloy and can be estimated by Aziz function~\cite{Aziz1982} given by: $k_v = (k + V/V_D)/(1 + V/V_D)$. When our phase-field $k_v$ predictions are compared to this relationship, we obtain a solute trapping map (Fig.~\ref{fig_aziz}) where the quantity $((k-k_v)/(1-k_v))$ is plotted against $V$ and a linear regression model is fit to the data to obtain the solute trapping gradient $1/V_D = W/D$. On average, the magnitude of solute trapping, i.e. $(k_v - k)$, increases with increasing $V$ with maximum absolute uncertainty of 0.3 is predicted by our simulations (Fig.~\ref{fig_aziz}).

We numerically optimize an objective function:
\begin{equation}
\argmin_{V_D} \frac{1}{N} \sum_{i = 1}^{N} (k_v^i - k_v^i(V_D))^2 \quad\forall i\in\left\{ 1,\ldots,N\right\},
\end{equation}
to find the root for $V_D$, which is estimated to be on the order of 6.4 m s$^{-1}$. When simulations were performed with the same phase-field model but with a limited input ($i<10$) of $V$, the $V_D$ obtained was on the order of 0.5 m s$^{-1}$~\cite{supriyo20173d}. Clearly, the size and distribution of the model inputs affect the calculation of interfacial properties, i.e., $V_D$. Experiments~\cite{Tian2017} determined the $V_D$ to be on the order of 9.0 m s$^{-1}$ in Ni-Nb alloys. We wish to note that a generalization of the Gulliver-Scheil model~\cite{rappazbook} and Aziz model~\cite{Aziz1982} could also indicate the influence of the inputs -- $V$, $k$, and $D$ -- on $k_v$ for rapid planar solidification at a much-reduced complexity compared to phase-field simulations. A comparison of the uncertainty treatments based on the analytical and phase-field methods is an interesting future research direction.

Frequency analysis of $k_v$ distribution was performed using CDFs and PDFs. Through an analysis of CDF, it was found that the estimated CDF that fits closely to the empirical CDF was Weibull and beta, with their differences being negligible. Representation of our microsegregation dataset using such a distribution not only reduces the amount of information conveyed by the dataset, but also describes the trends in the microsegregation patterns that develop during LPBF solidification. The uncertainty of such a distribution can be determined using an entropy function $\mathbb{H}(k_v)$, which uses the probability $p$ that is discretized into $K$ states to represent the $k_v$ distribution. The estimated $\mathbb{H}(k_v)$ can then be used to calculate the Kullback-Leibler divergence~\cite{murphy2012} to assess the dissimilarity between two probability distributions, say $k_{v_1}$ and $k_{v_2}$, using
\begin{eqnarray}\label{eq_kl}
\mathbb{H}(k_v) & \triangleq & - \sum_{t=1}^{K} p(k_v=t) \log p(k_v=t) \; \text{and} \nonumber\\ 
\mathbb{KL} (k_{v_1}||k_{v_2}) & \triangleq & \sum_{t=1}^{K} k_{v_1}^{t} \log\frac{k_{v_1}^{t}}{k_{v_2}^{t}}.
\end{eqnarray} 
In the literature, a Weibull distribution best described the macro-segregation behavior in the low-velocity solidification experiments and simulations~\cite{voller2014,fezi2016}. A lognormal distribution was shown to fit best the grain size distribution (a consequence of microsegregation) that resulted during additive manufacturing experiments on Ti-6Al-4V~\cite{Gregory2015}. In our calculations of a Ni-Nb alloy, estimated CDFs such as lognormal, normal and gamma are found to be either overestimating or underestimating in the tail, middle, or the top regions of the empirical CDF (Fig.~\ref{fig_cdf}). When experimental measurements will be available, our analysis will be used to validate our simulations.

\section{Summary and Outlook}\label{sec_summary}
Our work is summarized as follows:
\begin{itemize}
\item Forward propagation of process and alloy parameter uncertainties through mesoscale phase-field simulations has been conducted to study the LPBF solidification process of a Ni-Nb alloy. 
\item Uncertainty in model parameters leads to significant variability in microstructural features. The sample size and distribution of the model parameters severely affect the QoI distribution in the LPBF solidification regime.
\item Following the correlation measurements, we recommend that some less important parameters, $G$ and $\Gamma$, to be fixed at certain specific values (e.g., mean), and the variability in $V$, $D$, and $k$ needs to be considered.
\item A Gaussian process surrogate model satisfactorily approximates the phase-field predictions and hence can be used as a substitute for the phase-field method.
\item A frequency analysis identifies and quantifies the microsegregation distribution in the LPBF regime that affects the built material properties at the macroscale.
\end{itemize}

Additive manufacturing has the potential to be the technology for the future, and the quality control in LPBF can be achieved through variation control of the QoI~\cite{King2015,ghosh_review}. To achieve this, a better understanding of the uncertainty quantification of the LPBF process is essential. Our model and approach, in its current form, can be used as a first step towards microstructural engineering by providing users with computationally inexpensive predictions to explore the effects of model inputs on LPBF microstructures. The same approach could potentially be applied to investigate the variability on other key microstructural features, such as the dendrite arm spacing and the misorientation between dendrites. Future work will expand the proposed approach to an inverse problem on reducing the uncertainty by calibrating the model parameters using experimental measurements. The surrogate model can then be used to replace the expensive multi-scale, multi-physics phase-field model to meet the requirements of LPBF microstructures rapidly. Our ultimate goal is to build an efficient yet inexpensive framework for quantification of the datasets and distributions of the process and microstructure features in the additive manufacturing solidification regime.
\section*{Acknowledgements}
Authors would like to thank the support of the National Science Foundation Grant Nos. CMMI-1534534, CMMI-1663130, and DGE-1545403. RA and DA also acknowledge the support of ARL through Grant No. W911NF-132-0018. Portions of this work were also supported by an Early Stage Innovations grant from  NASA's  Space  Technology  Research
Grants Program, Grant No. NNX15AD71G. High-throughput phase-field simulations were carried out at the Ada and Terra Texas A\&M University supercomputing facilities.

\appendix
\section{Gaussian Process Surrogate Model}\label{appendix_surrogate}
We construct a Gaussian process (GP) surrogete model that has been detailed in Ref.~\cite{rasmussen2006,conti2010,mahmoudi2018}. A GP model assumes that given a finite input vector of $n$ variables, $\bm{X} = \lbrace \bm{x}_{1},\ldots,\bm{x}_{n} \rbrace$, the model outputs $\bm{Y} = \lbrace \bm{y}_{1},\ldots,\bm{y}_{n} \rbrace$ follow a \emph{q}-dimensional Gaussian process. Therefore, their joint probability distribution becomes a matrix normal distribution
\begin{equation}\label{eq_matrix-normal-dist}
\bm{Y} \mid \bm{\Phi} \sim \mathcal{MN}_{n,q}\left(\bm{m},\bm{C}\right),
\end{equation}
where $\bm{m}$ is the mean matrix and $\bm{C}$ is the covariance matrix, which are fully defined by a set of hyper-parameters $\bm{\Phi}$.
The multivariate GP is expressed as
\begin{equation}
\bm{Y} \mid \bm{\Phi} \sim \mathcal{GP}_{q}\left(m\left(\cdot\right),c\left(\cdot,\cdot\right)\bm{\Sigma}\right),
\end{equation}\label{eq_mvgp}
where $m\left(\cdot\right)$ is the mean function, $c\left(\cdot,\cdot\right)$ is a correlation function and $\bm{\Sigma}$ is a correlation matrix. Mean and correlation functions evaluated at $i=j=N$ number of ($\bm{X}$, $\bm{Y}$) training  points are defined as:
\begin{align}
m\left(\bm{x}\right) & =\bm{B}^{\top} h\left(\bm{x}\right)\, \text{and} \label{eq_mean} \\
c\left(\bm{x}_{i}, \bm{x}_{j}\right) & = \exp\left[-\left( \bm{x}_{i}-\bm{x}_{j}\right)^{\top} R\left( \bm{x}_{i}-\bm{x}_{j}\right)\right], \label{eq_corr}
\end{align}
where $h$ are regression functions that maps the input space to $m$ basis functions with regression coefficients $\bm{B}$, $\bm{x}_i = \lbrace x_{i,1},\cdots,x_{i,n} \rbrace_{i=1}^{N}$ the $i$-th training point, and $R={\rm diag}\left(\bm{r}\right)$ a diagonal matrix of positive roughness parameters with $\bm{r}=\lbrace r_{1},\ldots,r_{n} \rbrace$. $\bm{r}$ signifies the smoothness of the function. With the help of these functions, the GP model is fully defined as
\begin{equation}\label{eq_hyperparameters}
\bm{\Phi}=\left\{ \bm{B},\bm{\Sigma},\bm{r}\right\}. 
\end{equation}

It is described in Ref.~\cite{conti2010} that the conditional posterior distribution of $\bm{Y}$ given $\bm{r}$ (after integrating out $\bm{B}$ and $\bm{\Sigma}$) is a multivariate \emph{q}-dimensional T Process $f\left(\cdot\right)$ such that the resultant probability density follows a matrix-variate T distribution with a degrees of freedom $(N-m)$:
\begin{equation}
f\left(\cdot\right) \mid \bm{X},\bm{Y},\bm{r} \sim \mathcal{TP}_{q} \left(\hat{m}\left(\cdot\right),\hat{c}\left(\cdot,\cdot\right)\hat{\bm{\Sigma}},N-m\right).\label{eq:t-process}
\end{equation}
The $\hat{m}$ and $\hat{c}$ functions are defined by
\begin{align}
\hat{m}\left(\bm{x}\right) & =\hat{\bm{B}}^{\top}h\left(\bm{x}\right)+\left(\bm{Y}-\bm{H}\hat{\bm{B}}\right)^{\top}\bm{A}^{-1}\bm{t}\left(\bm{x}\right) \, \text{and}\label{eq:m-star}\\
\hat{c}\left(\bm{x}_{i}, \bm{x}_{j}\right) & =c\left(\bm{x}_{i},\bm{x}_{j}\right)-\bm{t}^{\top}(\bm{x}_{i})\bm{A}^{-1}\bm{t} (\bm{x}_{j})\nonumber \\
 & \hphantom{=}+\left[h (\bm{x}_{i})-\bm{H}^{\top}\bm{A}^{-1}\bm{t} (\bm{x}_{i})\right]^{\top}\left(\bm{H}^{\top}\bm{A}^{-1}\bm{H}\right)^{-1}\left[h (\bm{x}_{j})-\bm{H}^{\top}\bm{A}^{-1}\bm{t} (\bm{x}_{j})\right],\label{eq:c-star}
\end{align}
respectively. $\bm{H}^{\top} =\left[h (\bm{x}_{1}),\ldots,h (\bm{x}_{N})\right]$, $\bm{A} = c(\bm{x}_{i}, \bm{x}_{j})$, $\bm{t}^{\top} (\bm{x}_{i})$ = $\left[c\left( \bm{x}_{i}, \bm{x}_{1} \right),\ldots, c\left( \bm{x}_{i}, \bm{x}_{N} \right)\right]$, $\hat{\bm{B}} = \left(\bm{H}^{\top}\bm{A}^{-1}\bm{H}\right)^{-1}\bm{H}^{\top}\bm{A}^{-1}\bm{Y}$, and $\hat{\bm{\Sigma}} =  \left(N-m\right)^{-1}\left(\bm{Y}-\bm{H}\hat{\bm{B}}\right)^{\top}\bm{A}^{-1}\left(\bm{Y}-\bm{H}\hat{\bm{B}}\right)$. Equations~(\ref{eq:t-process}) to (\ref{eq:c-star}) are used as a surrogate for the phase-field model.

\begin{figure}[h]
\centering
\includegraphics[width=0.5\textwidth]{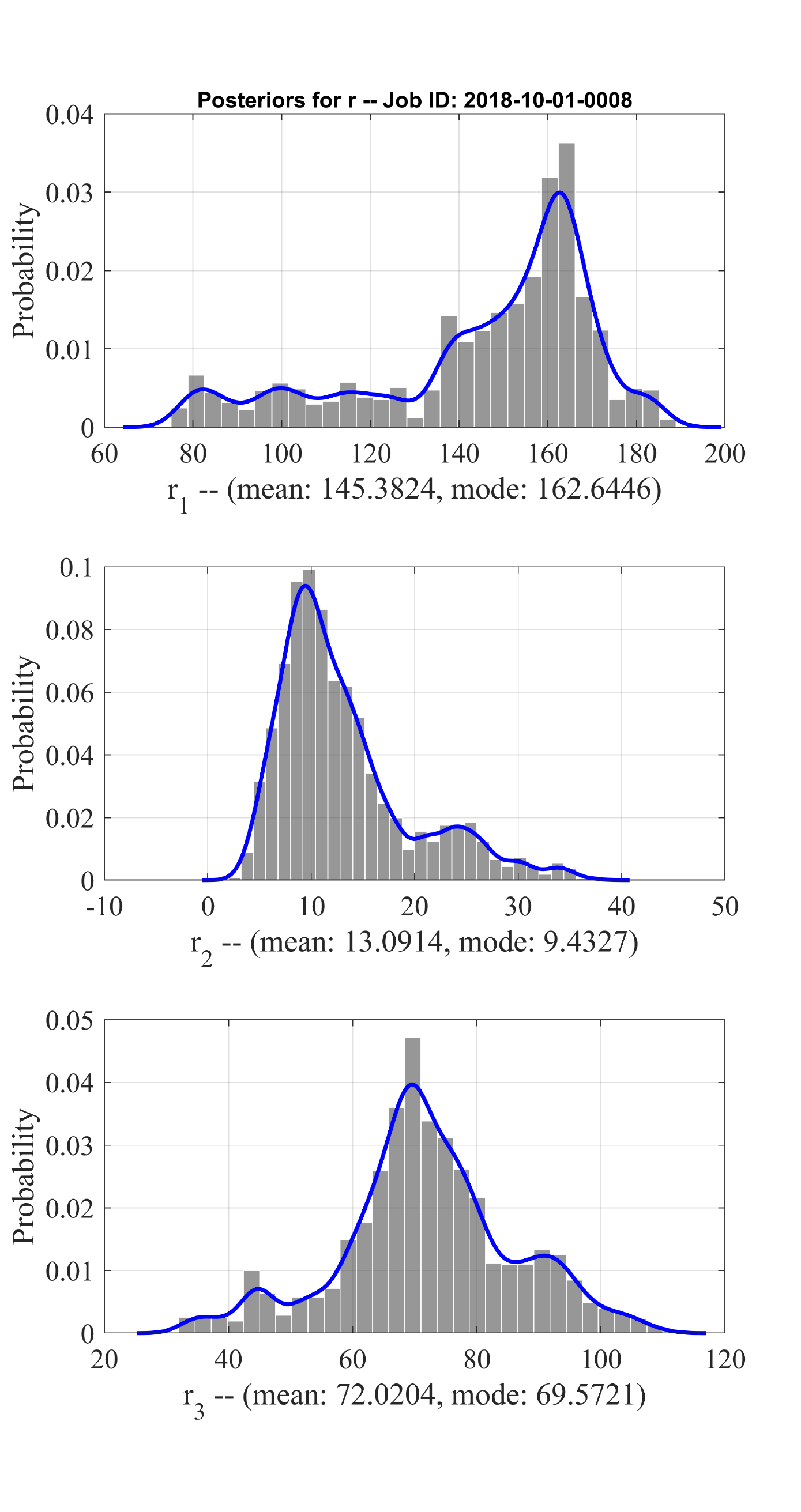}
\caption{Histograms and kernel density estimates of the posterior distributions for the roughness parameters $\bm{r}$ for the surrogate model with 3 input parameters: $\bm{X} = \lbrace D, V, k\rbrace$.}\label{fig:rHat}
\end{figure}

To estimate the roughness parameter $\bm{r}$, the Bayesian approach is used given a positive, log-logistic prior distribution. Using the single-component Metropolis-Hastings algorithm, the posterior distributions of $\bm{r}$ were generated after 20,000 iterations with 25\% burn-in period and thinning every fifth sample. Two separate surrogate models were trained: one surrogate model that uses 3 inputs $\bm{X} = \lbrace V, D, k \rbrace$ and another surrogate model that uses 5 inputs: $\bm{X} = \lbrace G, V, D, \Gamma, k \rbrace $. Figure~\ref{fig:rHat} shows the histograms and kernel density estimates of the posterior distributions for these inputs. We observe that the posteriors are unimodal, the modes of which were used as the estimates for $\bm{r}$. At this stage, the surrogate model in Eq.~(\ref{eq:t-process}) is fully defined, and the output of the phase-field model at any given $\bm{x}_i$ can be estimated using Eq.~(\ref{eq:m-star}). A confidence interval for this estimate can also be determined using Eq.~(\ref{eq:c-star}).

When all the components of $\bm{\Phi}$ (Eq.~(\ref{eq_hyperparameters})) are estimated, we assess the performance of the GP surrogate model through a \emph{p-fold cross validation} (CV) to ensure that the predictions obtained using the surrogate model are close to those obtained using the phase-field model. Our CV procedure separates the training dataset $\left(\bm{X},\bm{Y}\right)$ into \emph{p} disjoint partitions, among which $(p-1)$ of these partitions are used to train the surrogate model, and then the predictions are made on the left-out partition using Eq.~(\ref{eq:m-star}). These predictions are then compared with the phase-field predictions. This process is iterated \emph{p} times, such that at each iteration, a different partition is left out and after \emph{p} iterations all partitions have been left out only once. The comparison between surrogate and phase-field predictions is given in Sec.~\ref{sec_surrogate}.
\section*{References}
\bibliography{papers}

\begin{thebibliography}{10}
\expandafter\ifx\csname url\endcsname\relax
  \def\url#1{\texttt{#1}}\fi
\expandafter\ifx\csname urlprefix\endcsname\relax\def\urlprefix{URL }\fi
\expandafter\ifx\csname href\endcsname\relax
  \def\href#1#2{#2} \def\path#1{#1}\fi

\bibitem{frazier2014}
W.~E. Frazier, Metal {A}dditive {M}anufacturing: {A} {R}eview, Journal of
  Materials Engineering and Performance 23~(6) (2014) 1917--1928.

\bibitem{Herzog2016}
D.~Herzog, V.~Seyda, E.~Wycisk, C.~Emmelmann, Additive manufacturing of metals,
  Acta Materialia 117 (2016) 371 -- 392.

\bibitem{supriyo2017}
S.~Ghosh, L.~Ma, N.~Ofori-Opoku, J.~E. Guyer, On the primary spacing and
  microsegregation of cellular dendrites in laser deposited {N}i-{N}b alloys,
  Modelling and simulation in materials science and engineering 25~(6) (2017)
  065002.

\bibitem{Khairallah2016}
S.~A. Khairallah, A.~T. Anderson, A.~Rubenchik, W.~E. King, Laser powder-bed
  fusion additive manufacturing: {P}hysics of complex melt flow and formation
  mechanisms of pores, spatter, and denudation zones, Acta {M}aterialia 108
  (2016) 36 -- 45.

\bibitem{King2014}
W.~E. King, H.~D. Barth, V.~M. Castillo, G.~F. Gallegos, J.~W. Gibbs, D.~E.
  Hahn, C.~Kamath, A.~M. Rubenchik, Observation of keyhole-mode laser melting
  in laser powder-bed fusion additive manufacturing, Journal of {M}aterials
  {P}rocessing {T}echnology 214~(12) (2014) 2915 -- 2925.

\bibitem{King2015}
W.~E. King, A.~T. Anderson, R.~M. Ferencz, N.~E. Hodge, C.~Kamath, S.~A.
  Khairallah, A.~M. Rubenchik, Laser powder bed fusion additive manufacturing
  of metals; physics, computational, and materials challenges, Applied
  {P}hysics {R}eviews 2~(4) (2015) 041304.

\bibitem{boettinger2002}
W.~J. Boettinger, J.~A. Warren, C.~Beckermann, A.~Karma, Phase-field simulation
  of solidification, Annu. Rev. Mater. Res. 32 (2002) 163--194.

\bibitem{chen2002}
L.~Q. Chen, Phase-field models for microstructure evolution, Annu. Rev. Mater.
  Res. 32 (2002) 113--140.

\bibitem{steinbach2009}
I.~Steinbach, Phase-field models in materials science, Modelling and Simulation
  in Materials Science and Engineering 17 (2009) 073001.

\bibitem{attari2018}
V.~Attari, S.~Ghosh, T.~Duong, R.~Arroyave, On the interfacial phase growth and
  vacancy evolution during accelerated electromigration in {C}u/{S}n/{C}u
  microjoints, Acta Materialia 160 (2018) 185 -- 198.

\bibitem{ghosh2017_spinodal}
S.~Ghosh, A.~Mukherjee, T.~Abinandanan, S.~Bose, Particles with selective
  wetting affect spinodal decomposition microstructures, Physical Chemistry
  Chemical Physics 19~(23) (2017) 15424--15432.

\bibitem{tamoghna2018}
T.~Chakrabarti, S.~Manna, Zener pinning through coherent precipitate: {A}
  phase-field study, Computational Materials Science 154 (2018) 84 -- 90.

\bibitem{allaire2010}
D.~Allaire, K.~Willcox, Surrogate modeling for uncertainty assessment with
  application to aviation environmental system models, AIAA journal 48~(8)
  (2010) 1791--1803.

\bibitem{Franco2017_sr}
B.~Franco, J.~Ma, B.~Loveall, G.~Tapia, K.~Karayagiz, J.~Liu, A.~Elwany,
  R.~Arroyave, I.~Karaman, A sensory material approach for reducing variability
  in additively manufactured metal parts, Scientific Reports 7~(1) (2017) 3604.

\bibitem{ghosh_review}
S.~Ghosh, Predictive modeling of solidification during laser additive
  manufacturing of nickel superalloys: recent developments, future directions,
  Materials Research Express 5~(1) (2018) 012001.

\bibitem{debroy2015}
H.~Wei, J.~Mazumder, T.~DebRoy, Evolution of solidification texture during
  additive manufacturing, Scientific reports 5 (2015) 16446.

\bibitem{mahadevan2018}
P.~Nath, Z.~Hu, S.~Mahadevan, Modeling and uncertainty quantification of
  material properties in additive manufacturing, in: 2018 AIAA
  Non-Deterministic Approaches Conference, 2018, p. 0923.

\bibitem{mahadevan2017}
Z.~Hu, S.~Mahadevan, Uncertainty quantification and management in additive
  manufacturing: current status, needs, and opportunities, The International
  Journal of Advanced Manufacturing Technology 93~(5-8) (2017) 2855--2874.

\bibitem{brandon2016}
F.~Lopez, P.~Witherell, B.~Lane, Identifying uncertainty in laser powder bed
  fusion additive manufacturing models, Journal of Mechanical Design 138~(11)
  (2016) 114502.

\bibitem{claire2018}
C.~Bruna-Rosso, A.~G. Demir, M.~Vedani, B.~Previtali, Global sensitivity
  analyses of a selective laser melting finite element model: influential
  parameters identification, The International Journal of Advanced
  Manufacturing Technology\href {http://dx.doi.org/10.1007/s00170-018-2531-7}
  {\path{doi:10.1007/s00170-018-2531-7}}.

\bibitem{zaeem2013}
M.~A. Zaeem, H.~Yin, S.~D. Felicelli, Modeling dendritic solidification of
  {A}l--3\% {C}u using cellular automaton and phase-field methods, Applied
  Mathematical Modelling 37~(5) (2013) 3495--3503.

\bibitem{rappaz2016}
M.~Rappaz, Modeling and characterization of grain structures and defects in
  solidification, Current Opinion in Solid State and Materials Science 20~(1)
  (2016) 37--45.

\bibitem{supriyo20173d}
S.~Ghosh, N.~Ofori-Opoku, J.~E. Guyer, Simulation and analysis of $\gamma$-{N}i
  cellular growth during laser powder deposition of {N}i-based superalloys,
  Computational Materials Science 144 (2018) 256--264.

\bibitem{kundin2015}
J.~Kundin, L.~Mushongera, H.~Emmerich, Phase-field modeling of microstructure
  formation during rapid solidification in {I}nconel 718 superalloy, Acta
  Materialia 95 (2015) 343 -- 356.

\bibitem{steinbach2012}
L.~Zhang, I.~Steinbach, Phase-field model with finite interface dissipation:
  {E}xtension to multi-component multi-phase alloys, Acta Materialia 60~(6-7)
  (2012) 2702--2710.

\bibitem{Echebarria2004}
B.~Echebarria, R.~Folch, A.~Karma, M.~Plapp, Quantitative phase-field model of
  alloy solidification, Physical {R}eview {E} 70~(6) (2004) 061604.

\bibitem{Karma2001}
A.~Karma, Phase-field formulation for quantitative modeling of alloy
  solidification, Physical {R}eview {L}etters 87 (2001) 115701.

\bibitem{provatas2011}
N.~Provatas, K.~Elder, Phase-field methods in materials science and
  engineering, John Wiley \& Sons, 2011.

\bibitem{baker1992}
H.~Baker, ASM handbook: Alloy phase diagrams, no. v. 3 in ASM Handbook: Alloy
  Phase Diagrams, ASM International, 1992.

\bibitem{knorovsky1989}
G.~A. Knorovsky, M.~J. Cieslak, T.~J. Headley, A.~D. Romig, W.~F. Hammetter,
  {INCONEL} 718: A solidification diagram, Metallurgical Transactions A 20~(10)
  (1989) 2149--2158.

\bibitem{Comsol}
{COMSOL}, Multiphysics Reference Guide for COMSOL 4.2 (2011).

\bibitem{trapp2017}
J.~Trapp, A.~M. Rubenchik, G.~Guss, M.~J. Matthews, In situ absorptivity
  measurements of metallic powders during laser powder-bed fusion additive
  manufacturing, Applied Materials Today 9 (2017) 341--349.

\bibitem{karayagiz2018}
K.~Karayagiz, A.~Elwany, G.~Tapia, B.~Franco, L.~Johnson, J.~Ma, I.~Karaman,
  R.~Arroyave, Numerical and experimental analysis of heat distribution in the
  laser powder bed fusion of {T}i-6{A}l-4{V}, IISE Transactions~(just-accepted)
  (2018) 1--44.

\bibitem{bolten1984}
A.~Block-Bolten, T.~W. Eagar, Metal vaporization from weld pools, Metallurgical
  Transactions B 15~(3) (1984) 461--469.

\bibitem{asmnickel}
J.~Davis, A.~Committee, Nickel, Cobalt, and Their Alloys, ASM specialty
  handbook, ASM International, 2000.

\bibitem{agarwal2004}
D.~C. Agarwal, Nickel and Nickel Alloys, Wiley-Blackwell, 2004, Ch.~7, pp.
  217--270.

\bibitem{ladani2017}
L.~Ladani, J.~Romano, W.~Brindley, S.~Burlatsky, Effective liquid conductivity
  for improved simulation of thermal transport in laser beam melting powder bed
  technology, Additive Manufacturing 14 (2017) 13--23.

\bibitem{roberts2009}
I.~A. Roberts, C.~J. Wang, R.~Esterlein, M.~Stanford, D.~J. Mynors, A
  three-dimensional finite element analysis of the temperature field during
  laser melting of metal powders in additive layer manufacturing, International
  Journal of Machine Tools and Manufacture 49~(12) (2009) 916--923.

\bibitem{raghavan2016}
N.~Raghavan, R.~Dehoff, S.~Pannala, S.~Simunovic, M.~Kirka, J.~Turner,
  N.~Carlson, S.~S. Babu, Numerical modeling of heat-transfer and the influence
  of process parameters on tailoring the grain morphology of {IN718} in
  electron beam additive manufacturing, Acta Materialia 112 (2016) 303--314.

\bibitem{mahmoudi2018}
M.~Mahmoudi, G.~Tapia, K.~Karayagiz, B.~Franco, J.~Ma, R.~Arroyave, I.~Karaman,
  A.~Elwany, Multivariate calibration and experimental validation of a 3d
  finite element thermal model for laser powder bed fusion metal additive
  manufacturing, Integrating Materials and Manufacturing Innovation 7~(3)
  (2018) 116--135.

\bibitem{lee2010}
L.~Yuan, P.~D. Lee, Dendritic solidification under natural and forced
  convection in binary alloys: {2D} versus {3D} simulation, Modelling and
  Simulation in Materials Science and Engineering 18~(5) (2010) 055008.

\bibitem{olsson2003latin}
A.~Olsson, G.~Sandberg, O.~Dahlblom, On latin hypercube sampling for structural
  reliability analysis, Structural safety 25~(1) (2003) 47--68.

\bibitem{montgomery2014}
D.~Montgomery, G.~Runger, Applied Statistics and Probability for Engineers,
  Wiley, 2014.

\bibitem{devore1987}
J.~Devore, Probability and Statistics for Engineering and the Sciences,
  Brooks/Cole, 1987.

\bibitem{kennedy2001}
M.~C. Kennedy, A.~O'Hagan, Bayesian calibration of computer models, Journal of
  the Royal Statistical Society: Series B (Statistical Methodology) 63~(3)
  (2001) 425--464.

\bibitem{Matlab}
MATLAB, Version R2018a, The MathWorks Inc., Natick, Massachusetts, 2018.

\bibitem{rappazbook}
M.~Rappaz, J.~A. Dantzig, Solidification, Engineering sciences, EFPL Press,
  2009.

\bibitem{Mullins1964}
W.~W. Mullins, R.~F. Sekerka, Stability of a planar interface during
  solidification of a dilute binary alloy, Journal of Applied Physics 35~(2)
  (1964) 444--451.

\bibitem{ohagan2013}
A.~O'Hagan, Polynomial chaos: A tutorial and critique from a statistician's
  perspective, SIAM/ASA J. Uncertainty Quantification 20 (2013) 1--20.

\bibitem{Mahmoudi2018MVCalibration}
M.~Mahmoudi, G.~Tapia, {Multivariate Statistical Calibration of Computer
  Models}, \url{https://github.com/mahmoudi-tapia/MVcalibration} (2018).

\bibitem{mahadevan2012}
S.~Sankararaman, S.~Mahadevan, Likelihood-based approach to multidisciplinary
  analysis under uncertainty, Journal of Mechanical Design 134~(3) (2012)
  031008.

\bibitem{Trevor2017}
T.~Keller, G.~Lindwall, S.~Ghosh, L.~Ma, B.~Lane, F.~Zhang, U.~R. Kattner,
  E.~A. Lass, J.~C. Heigel, Y.~Idell, M.~E. Williams, A.~J. Allen, J.~E. Guyer,
  L.~E. Levine, Application of {F}inite {E}lement, {P}hase-field, and
  {CALPHAD}-based {M}ethods to {A}dditive {M}anufacturing of {N}i-based
  {S}uperalloys, Acta {M}aterialia 139 (2017) 244--253.

\bibitem{ranadip2017}
R.~Acharya, J.~A. Sharon, A.~Staroselsky, Prediction of microstructure in laser
  powder bed fusion process, Acta Materialia 124 (2017) 360--371.

\bibitem{ghosh2018single}
S.~Ghosh, L.~Ma, L.~E. Levine, R.~E. Ricker, M.~R. Stoudt, J.~C. Heigel, J.~E.
  Guyer, Single-track melt-pool measurements and microstructures in inconel
  625, JOM (2018) 1--6.

\bibitem{Ghosh2018}
S.~Ghosh, M.~R. Stoudt, L.~E. Levine, J.~E. Guyer, Formation of {N}b-rich
  droplets in laser deposited {N}i-matrix microstructures, Scripta Materialia
  146 (2018) 36--40.

\bibitem{vrancken2014}
B.~Vrancken, L.~Thijs, J.-P. Kruth, J.~Van~Humbeeck, Microstructure and
  mechanical properties of a novel $\beta$ titanium metallic composite by
  selective laser melting, Acta Materialia 68 (2014) 150--158.

\bibitem{tao2019}
P.~Tao, H.~Li, B.~Huang, Q.~Hu, S.~Gong, Q.~Xu, The crystal growth,
  intercellular spacing and microsegregation of selective laser melted
  {I}nconel 718 superalloy, Vacuum 159 (2019) 382 -- 390.

\bibitem{Boettinger1999}
W.~J. Boettinger, J.~A. Warren, Simulation of the cell to plane front
  transition during directional solidification at high velocity, Journal of
  Crystal Growth 200 (1999) 583 -- 591.

\bibitem{li2016_sobol}
C.~Li, S.~Mahadevan, An efficient modularized sample-based method to estimate
  the first-order sobol' index, Reliability Engineering \& System Safety 153
  (2016) 110--121.

\bibitem{tapia2018}
G.~Tapia, W.~E. King, R.~Arroyave, L.~Johnson, I.~Karaman, A.~Elwany,
  Validation of a laser-based powder bed fusion thermal model via uncertainty
  propagation and generalized polynomial chaos expansions, Journal of
  Manufacturing Science and Engineering\href
  {http://dx.doi.org/10.1115/1.4041179} {\path{doi:10.1115/1.4041179}}.

\bibitem{aiken1991}
L.~Aiken, S.~West, R.~Reno, Multiple Regression: Testing and Interpreting
  Interactions, SAGE Publications, 1991.

\bibitem{popova2017}
E.~Popova, T.~M. Rodgers, X.~Gong, A.~Cecen, J.~D. Madison, S.~R. Kalidindi,
  Process-structure linkages using a data science approach: application to
  simulated additive manufacturing data, Integrating Materials and
  Manufacturing Innovation 6~(1) (2017) 54--68.

\bibitem{jung2019}
J.~Jung, J.~I. Yoon, H.~K. Park, J.~Y. Kim, H.~S. Kim, An efficient machine
  learning approach to establish structure-property linkages, Computational
  Materials Science 156 (2019) 17 -- 25.

\bibitem{kennedy2000}
M.~C. Kennedy, A.~O'Hagan, Predicting the output from a complex computer code
  when fast approximations are available, Biometrika 87~(1) (2000) 1--13.

\bibitem{tuo2014}
R.~Tuo, C.~J. Wu, D.~Yu, Surrogate modeling of computer experiments with
  different mesh densities, Technometrics 56~(3) (2014) 372--380.

\bibitem{huan2014gradient}
X.~Huan, Y.~Marzouk, Gradient-based stochastic optimization methods in
  {B}ayesian experimental design, International Journal for Uncertainty
  Quantification 4~(6).

\bibitem{ghoreishi2018multi}
S.~F. Ghoreishi, A.~Molkeri, A.~Srivastava, R.~Arroyave, D.~Allaire,
  Multi-information source fusion and optimization to realize {ICME}:
  {A}pplication to dual-phase materials, Journal of Mechanical Design 140~(11)
  (2018) 111409.

\bibitem{mullis2010}
A.~M. Mullins, J.~Rosam, P.~K. Jimack, Solute trapping and the effects of
  anti-trapping currents on phase-field models of coupled thermo-solutal
  solidification, Journal of Crystal Growth 312~(11) (2010) 1891 -- 1897.

\bibitem{ohno2009}
M.~Ohno, K.~Matsuura, Quantitative phase-field modeling for dilute alloy
  solidification involving diffusion in the solid, Physical Review E 79~(3)
  (2009) 031603.

\bibitem{Aziz1982}
M.~J. Aziz, Model for solute redistribution during rapid solidification,
  Journal of Applied Physics 53~(2) (1982) 1158--1168.

\bibitem{Tian2017}
Y.~Tian, J.~Mu\={n}iz-Lerma, M.~Brochu, Nickel-based superalloy microstructure
  obtained by pulsed laser powder bed fusion, Materials Characterization 131
  (2017) 306 -- 315.

\bibitem{murphy2012}
K.~Murphy, F.~Bach, Machine {L}earning: {A} {P}robabilistic {P}erspective,
  Adaptive Computation and Machi, MIT Press, 2012.

\bibitem{voller2014}
V.~Voller, I.~Vu{\v{s}}anovi{\'c}, Frequency analysis of macrosegregation
  measurements and simulations, International Journal of Heat and Mass Transfer
  79 (2014) 468--471.

\bibitem{fezi2016}
K.~Fezi, A.~Plotkowski, M.~J. Krane, A metric for the quantification of
  macrosegregation during alloy solidification, Metallurgical and Materials
  Transactions A 47~(6) (2016) 2940--2951.

\bibitem{Gregory2015}
G.~Loughnane, A framework for uncertainty quantification in microstructural
  characterization with application to additive manufacturing of
  {T}i-6{A}l-4{V}, Ph.D. thesis, Wright State University (2015).

\bibitem{rasmussen2006}
C.~Rasmussen, C.~Williams, Gaussian Processes for Machine Learning, Adaptative
  computation and machine learning series, MIT Press, 2006.

\bibitem{conti2010}
S.~Conti, A.~O'Hagan, Bayesian emulation of complex multi-output and dynamic
  computer models, Journal of statistical planning and inference 140~(3) (2010)
  640--651.

\end{thebibliography}

\end{document}